\documentclass[11pt, oneside]{article}   	

\usepackage{showlabels}
\usepackage{jheppub}

\usepackage{amsmath}
\usepackage{amssymb}
\usepackage{amsfonts}
\usepackage{amsthm}
\usepackage{amsbsy}
\usepackage{array}

\usepackage{mathtools}
\usepackage{xcolor}

\usepackage[vcentermath]{youngtab}

\usepackage{tikz}
\usetikzlibrary{decorations.markings}

\theoremstyle{definition}

\theoremstyle{theorem}
\newtheorem{theorem}{Theorem}

\theoremstyle{definition}
\newtheorem{proposition}{Proposition}

\newtheorem*{examples}{Examples}

\makeatletter
\renewcommand*\env@matrix[1][\arraystretch]{%
  \edef\arraystretch{#1}%
  \hskip -\arraycolsep
  \let\@ifnextchar\new@ifnextchar
  \array{*\c@MaxMatrixCols c}}
\makeatother

\DeclarePairedDelimiter{\floor}{\lfloor}{\rfloor}

\newcommand{\pol}{s}

\newcommand{\spinor}{\mathbf{\frac{1}{2}}}

\newcommand{\beq}{\begin{equation}}
\newcommand{\eeq}{\end{equation}}

\newcommand{\up}{\uparrow}
\newcommand{\down}{\downarrow}

\newcommand{\al}[1]{\begin{align} #1\end{align}}

\newcommand{\crr}[1]{\left\langle #1 \right\rangle}

\newcommand*\Bell{\ensuremath{\boldsymbol\ell}}

\newcommand{\OO}{\mathcal{O}}

\newcommand{\RR}{\mathcal{R}}
\newcommand{\MM}{\mathcal{M}}

\newcommand{\Amp}{\mathcal{A}}
\newcommand{\PP}{\mathcal{P}}
\newcommand{\<}{\langle}
\renewcommand{\>}{\rangle}
\newcommand{\oo}{\infty}
\newcommand{\cO}{\mathcal{O}}
\newcommand{\Res}{\mathrm{Res}}
\newcommand{\p}[1]{\left(#1\right)}

\newcommand\nn\nonumber

\title{Counting Conformal Correlators}
\author{Petr Kravchuk$^{1}$ and David Simmons-Duffin$^{1,2}$}
\affiliation{${}^1$Walter Burke Institute for Theoretical Physics, Caltech, Pasadena, California 91125, USA \\
${}^2$School of Natural Sciences, Institute for Advanced Study, Princeton, New Jersey 08540, USA
}
\date{}							

\abstract{
We introduce simple group-theoretic techniques for classifying conformally-invariant tensor-structures. With them, we classify tensor structures of general $n$-point functions of non-conserved operators, and $n\geq 4$-point functions of general conserved 
 currents, with or without permutation symmetries, and in any spacetime dimension $d$. (The case $n=3$ for conserved operators will appear in subsequent work.) Our techniques are useful for bootstrap applications.
The rules we derive simultaneously count tensor structures for flat-space scattering amplitudes in $d+1$ dimensions.
}

\preprint{CALT-TH 2016-041}

\begin{document}

\maketitle

\section{Introduction}
\label{sec:introduction}

To apply conformal bootstrap techniques~\cite{Ferrara:1973yt,Polyakov:1974gs,Rattazzi:2008pe} to operators with spin, one must first understand the space of conformally-invariant tensor structures.
 This problem has been addressed previously for various types of operators in various dimensions~\cite{Sotkov:1976xe,mack1977,Osborn:1993cr,Weinberg:2010fx,Giombi:2011rz,Costa:2011mg,SimmonsDuffin:2012uy,Elkhidir:2014woa,Costa:2014rya,Iliesiu:2015qra}. However, no completely general construction or classification of tensor structures currently exists in the literature.

The approaches~\cite{Giombi:2011rz,Costa:2011mg,SimmonsDuffin:2012uy,Elkhidir:2014woa,Costa:2014rya,Iliesiu:2015qra} follow the strategy of defining basic conformally-invariant building blocks, and then multiplying them in all possible ways. While this strategy makes it easy to build conformally-invariant structures, it is not always convenient for bootstrap applications. This is because the building blocks satisfy nontrivial algebraic relations, which give rise to redundancies between structures built from them. As an example, of $201$ possible parity-even combinations of the building blocks of~\citep{Costa:2011mg} for the four-point function of identical spin-2 operators, only $97$ are linearly independent in 3 dimensions. It is possible in principle to find relations between the $201$ structures, and then choose a ``standard'' basis of $97$ independent structures. However, this task is technically complicated and one may wonder if this step can be omitted completely.

In this paper we discuss a different approach, which extends the formalism of~\cite{mack1977,Osborn:1993cr} to $n$-point functions. Based on the simple idea of ``gauge-fixing'' the conformal symmetry, our approach makes it possible to avoid the problem of algebraic relations completely in many cases. Furthermore, it applies uniformly to any operators in arbitrary representations of $SO(d)$, being essentially equivalent to invariant theory of orthogonal groups. 

The basic idea is simple. Consider a three-point function $\<\cO_1^{a_1}(x_1) \cO_2^{a_2}(x_2) \cO_3^{a_3}(x_3)\>$, where the operators $\cO_i$ transform in representations $\rho_i$ of the rotation group $SO(d)$, and $a_i$ are indices for those representations. Using conformal transformations, we can place the operators in a standard configuration, say $\<\cO_1^{a_1}(0)\cO_2^{a_2}(e)\cO_3^{a_3}(\oo)\>$, where $e$ is a unit vector. The correlator must then be invariant under the ``little group" for this configuration, which is the group $SO(d-1)$ of rotations that preserve the line through $0,e,\oo$. Such invariants are given by
\beq
\p{\Res^{SO(d)}_{SO(d-1)} \bigotimes_{i=1}^3 \rho_i}^{SO(d-1)},
\label{eq:3ptbasic}
\eeq
where $\Res^G_H$ denotes restriction from a representation of $G$ to a representation of $H\subseteq G$, and $(\rho)^H$ represents the $H$-invariant subspace of $\rho$ (i.e.\ the singlet sub-representations).

We generalize this argument in several directions: to arbitrary $n$-point functions, to incorporate permutation symmetries between identical operators, and most nontrivially to deal with conserved operators like currents $J^\mu$ and the stress-tensor $T^{\mu\nu}$. For three-point functions involving conserved operators, the conservation conditions become linear relations between tensor structures.\footnote{These linear relations have interesting structure that we explore in~\cite{KravchukFuture}.} However, for general $n$-point functions, conservation constraints become differential equations which are quite
complicated to analyze~\cite{Dymarsky:2013wla}. The conclusion of~\cite{Dymarsky:2013wla} is that such correlators can be parametrized by a smaller number of functions of the conformal invariants of $n$ points. For example, a parity-even four-point function of stress-tensors in 3d is parameterized by 5 scalar functions of conformal cross-ratios. We find a simple group-theoretic rule for counting these functions.

Besides simplicity, there are several motivations for characterizing the space of tensor structures in representation-theoretic language. Firstly, it is an obvious first step towards finding a general representation-theoretic formula for conformal blocks in $d>2$ dimensions. Many examples of conformal blocks (not to mention superconformal blocks) have been computed using a variety of techniques~\cite{DO1,DO2,DO3,Costa:2011dw,SimmonsDuffin:2012uy,Hogervorst:2013sma,Kos:2013tga,Hogervorst:2013kva,Kos:2014bka,Costa:2014rya,Iliesiu:2015akf,Penedones:2015aga,Costa:2016xah,Costa:2016hju,Echeverri:2016dun}, but no one technique has yet proved completely general and efficient. Secondly, similar language might be helpful in classifying superconformally-invariant tensor structures, about which much less is known.

Importantly for numerical applications, our approach allows us to construct the tensor structures explicitly. We work out the tensor structures of non-conserved operators in 3d as an example.

It is well known~\cite{Hofman:2008ar,Costa:2011mg,Costa:2014rya} that the number of conformally-invariant tensor structures for a correlator in $d$-dimensions is equal to the number of Lorentz and gauge invariant tensor structures for a flat space scattering amplitude in $d+1$-dimensions. We demonstrate this relation by interpreting our group-theoretic counting rules in the $S$-matrix context.

\section{Conformal correlators of long multiplets}
\label{sec:conformalcorrelators}

In this section we describe in detail the construction and counting of tensor structures for correlators of long conformal mulptiplets (local operators not constrained by differential equations).  

\subsection{Conformal invariance}
\label{sec:conformalinvariance}

Consider a Euclidean CFT${}_d$ on $\mathbb{R}^d$.\footnote{\label{foot:noteaboutinfinity}Actually, we work on the conformal compactification $S^d$ of $\mathbb{R}^d$, which means we can place operators at infinity. We will sometimes use the non-standard definition $\cO(\oo)\equiv \lim_{L\to \oo} L^{2\Delta_\cO} \cO(Le)$, with $e$ a fixed unit vector. The advantage of this definition is that we don't apply an inversion to $\cO$, so $\cO$ is treated more symmetrically with other operators in the correlator. The disadvantage is that the definition depends on $e$, so it breaks some rotational symmetries. However, in most of our computations these symmetries will already be broken by other operators in the correlator.} A conformally-invariant correlation function of $n$ primary operators $\OO_i^{a_i}(x_i)$ in representations $\rho_i$ of $SO(d)$ can be expressed as 
\beq
\crr{\OO_1^{a_1}(x_1)\ldots\OO_n^{a_n}(x_n)}=\sum_{I=1}^N \mathbb{Q}^{a_1\ldots a_n}_I(x_i) g^I(\mathbf{u}),\label{eq:genericcorr}
\eeq
where $g^I$ are scalar functions of the conformal invariants $\mathbf{u}$ of $n$ points, and the possible tensor structures $\mathbb{Q}_I^{a_1\ldots a_n}$ are constrained by conformal invariance. When some of the operators $\OO_i$ are identical, these structures are further constrained by symmetry with respect to permutations. When one or more of the operators is a conserved current, the correlator also satisfies nontrivial differential equations.

Let $SO_0(d+1,1)$ be the identity component of the conformal group. Conformal transformations $U\in SO_0(d+1,1)$ act on  primary operators as
\beq
	U\OO^a(x)U^{-1}=\Omega(x')^{\Delta}\rho^{a}{}_{b}(R(x')^{-1})\OO^b(x'),
\eeq
where 
\beq
	\Omega(x')R^{\mu}{}_\nu(x')=\frac{\partial x'^{\mu}}{\partial x^\nu},
\eeq
with $\Omega(x)>0$ and $R(x)\in SO(d)$. This leads to the following transformation of the correlator
\beq
	\crr{\OO_1^{a_1}(x_1)\ldots\OO_n^{a_n}(x_n)}=\left[\prod_{i=1}^n \Omega(x'_i)^{\Delta_i} \rho^{a_i}_i{}_{b_i}(R(x'_i)^{-1})\right]\crr{\OO_1^{b_1}(x_1')\ldots\OO_n^{b_n}(x_n')}.
\eeq

When some of the operators are fermionic, a small clarification is required. By construction, $R(x)$ is an element of $SO(d)$. However, it is the double cover $Spin(d)$ of $SO(d)$ that acts on a fermionic representation. One therefore must lift $R(x)\in SO(d)$ to some $\RR(x)\in Spin(d)$. A natural point of view is to assign $\RR(x)$ to an element $r$ of the double cover $Spin(d+1,1)$ of the conformal group $SO_0(d+1,1)$: first we assign $\RR(x)\equiv \mathrm{id}$ to the identity of $Spin(d+1,1)$ and then define $\RR$ on the rest of $Spin(d+1,1)$ by continuity. This is consistent because $Spin(d+1,1)$ is simply-connected. The invariance of correlation functions under the center of $Spin(d+1,1)$ is then simply the selection rule that the correlation function has to contain an even number of fermions.

To faciliate group-theoretic arguments, we write
\beq
g^{a_1\ldots a_n}(x_1,\ldots, x_n)=\crr{\OO^{a_1}_1(x_1)\ldots \OO^{a_n}_n(x_n)},
\eeq
and define the action of the conformal group on $g$ as follows. Let $r\in Spin(d+1,1)$ be a conformal transformation. It uniquely defines elements 
\beq
	\RR_r(x')\in Spin(d),\quad \Omega_r(x')>0,
\eeq
as described above. We define the action of $r$ on $g$ by
\beq
\label{eq:conformalgroupaction}
	(rg)^{a_1\ldots a_n}(x_i,\ldots,x_n)=\prod_{i=1}^n \Omega(x_i)^{-\Delta_i} \rho^{a_i}_i{}_{b_i}(\RR_r(x_i))g^{b_1\ldots b_n}(r^{-1}x_1,\ldots,r^{-1}x_n).
\eeq
With this definition, conformal invariane of the correlator is simply the statement that 
\beq
	rg=g.\label{eq:confinv}
\eeq

We will often parametrize operators by polarizations, $\OO(\pol,x)=\pol_a\OO^a(x)$. In this case $g$ becomes a function of $\pol_i$ as well as $x_i$, and the above action becomes
\beq
	(rg)(\pol_i,x_i)=\prod_{i=1}^n\Omega_r(x_i)^{-\Delta_i} g(\RR_r(x_i)^{-1}\pol_i, r^{-1}x_i),\label{eq:polarizationinvariance}
\eeq
where for simplicity of notation we implicitly assume that $\pol_i$ transforms in the dual representation $\rho_i^\vee$.

In a parity-preserving theory the above analysis should be extended to include reflections in $O(d)$. When fermions are present, one must specify a double cover $Pin(d)$ of $O(d)$ which will act on the spinor representations. In the following discussion this choice will be encapsulated in the representation theory of $Pin(d)$, and we therefore simply assume that a choice has been made which consistently defines an action of the disconnected conformal group on the correlators. In the following we will often refer to $SO(\cdot)$ or $O(\cdot)$ groups when we really mean their double covers if fermionic operators are involved. We hope that this will not cause confusion. 

\subsection{Conformal frame}
\label{sec:conformalframe}

Consider a four-point function of scalars,
\beq
	g(x_1,x_2,x_3,x_4)=\crr{\OO_1(x_1)\OO_2(x_2)\OO_3(x_3)\OO_4(x_4)}.
\eeq
It is well-known that $g(x_i)$ only depends on two variables, the cross-ratios $u$ and $v$,
\beq
\label{eq:crossratios}
	u=\frac{x_{12}^2x_{34}^2}{x_{13}^2x_{24}^2},\quad v=\frac{x_{14}^2x_{23}^2}{x_{13}^2x_{24}^2},
\eeq
where $x_{ij}=x_i-x_j$. The usual way to see this is to ``fix'' the conformal symmetry: choose a 2d half-plane $\alpha$, a vector $e\in \partial\alpha$, and use conformal symmetry to set $x_1=0$, $x_3=e$, and $x_4=\oo$. The remaining symmetry is just the $SO(d-1)$ of rotations that fix $e$. Using these, we can put $x_2$ in $\alpha$. Let us call the set of such configurations (when $x_1,x_3,x_4$ are fixed and $x_2\in\alpha$) a \textit{conformal frame}.

Since any configuration can be mapped by a conformal transformation to a conformal frame configuration, it's clear that the full correlator $g$ is uniquely fixed by its restriction $g_0$ to conformal frame configurations. These are parametrized by  two coordinates for the point $x_2$ in $\alpha$, which we can choose to be $u$ and $v$. 

With the coordinates $x_i$ brought to a conformal frame configuration $y_i$, $g_0$ must still be invariant under the ``little group." More precisely, let $St(y)\subset SO_0(d+1,1)$ be the group of conformal transformations that stabilize the $y_i$. Conformal invariance requires that for any $h\in St(y)$,
\beq
	g_0(y_i)=(hg_0)(y_i).\label{eq:confinvCF}
\eeq
For scalars this is automatic, since $St(y)$ is always a rotation group, and scalars are invariant under rotations. (For $y$ in the interior of conformal frame, $St(y)$ is the $SO(d-2)$ of rotations orthogonal to $\alpha$, and for $y$ on the boundary $St(y)$ is the $SO(d-1)$ that fixes $e$.) Assuming that~\eqref{eq:confinvCF} holds, we can consistently define the full correlator $g$ starting from $g_0$ by writing
\beq
	g(x_i)=(r_x g_0)(x_i),\label{eq:cfreconstruction}
\eeq
where $r_x$ is any conformal transformation such that $y_i=r^{-1}_x x_i$ is in the conformal frame.
The definition (\ref{eq:cfreconstruction}) doesn't depend on the choice of $r_x$ for the usual reason: any other $r'_x$ satisfies $r'_x=r_x h$ for some $h\in St(y)$, and this gives rise to the same $g(x_i)$ because of (\ref{eq:confinvCF}).

This approach clearly generalizes to $n$-point functions of operators in arbitrary $SO(d)$ representations  --- the only new ingredient is that the invariance \eqref{eq:confinvCF} under the stabilizer subgroup $St(y)$ is now a non-trivial constraint. Quite generally, the configuration space of $n$ points on the sphere splits into orbits under the action of the connected conformal group; we define the conformal frame to be a submanifold of the configuration space which intersects each orbit at precisely one point. Then all of the above works verbatim.

This is perhaps most striking for four-point functions in 3 dimensions. In this case, the stabilizer subgroup is generically the trivial $SO(3-2)=SO(1)$! So spinning four-point functions in 3d are almost no different from scalar ones. We return to this point in section~\ref{sec:3d4pt}.

Note that the above discussion showed that $St(y)$-invariance of $g_0$ is sufficient for $g$ to be well-defined, but not necessarily smooth. If we require $g$ to be smooth, we must impose more refined conditions for $g_0$ on the boundaries of the conformal frame. We discuss this point in appendix~\ref{sec:smoothness}. As we discuss in section~\ref{sec:3d4Majorana}, these conditions are important for formulating the bootstrap equations.

\subsection{$n$-point functions}
\label{sec:npointconformalframe}
Consider the general case of $n\geq 3$ points. For convenience, we define $m=\min(n,d+2)$.
To specify a conformal frame, we choose a flag of half-subspaces\footnote{If $m=d+2$, then $\alpha_d$ should be the full linear subspace instead of a half-space. This is because when we fix the position of the last operator, we can only use $SO(d+3-m)$, which is trivial in this case.} $\alpha_i$, $i=2,\ldots m-2$, such that
\al{
	&\dim\alpha_i=i,\nn\\
	&\partial \alpha_i=\overline\alpha_{i-1},\,\,i>2,\nn\\
	&\partial \alpha_2=\mathbb{R}e,
}
and $\overline \alpha_i$ is the linear subspace spanned by $\alpha_i$. We first put operators $1,2,3$ at $0,e,\infty$, as before. We then use the remaining $SO(d-i+3)$ to bring the $i$-th operator to lie in $\alpha_{i-2}$, for $i=4,\ldots,m$. If $n>m$, we have already used all the conformal symmetry to fix the positions of the first $m$ operators, and the remaining $n-m$ operators can be anywhere.

After this is done, a generic conformal frame configuration has stabilizer subgroup $SO(d+2-m)$. It follows that the conformally-invariant tensor structures are given by
\beq
\p{\mathrm{Res}_{SO(d+2-m)}^{SO(d)}\bigotimes_{i=1}^n\rho_i}^{SO(d+2-m)}.\label{eq:socounting}
\eeq 
Again, $\mathrm{Res}^G_H$ denotes the restriction of a representation of $G$ to a representation of $H\subseteq G$,\footnote{Because $\Res^G_H$ is a functor, we can restrict the representations before taking their tensor products. This sometimes simplifies calculations.} and $\rho_i$ are the $SO(d)$ representations of the $\OO_i$, and $(\rho)^H$ denotes the $H$-singlets in $\rho$.

This counting rule is consistent with the result of~\cite{Costa:2014rya}. For simplicity, consider three-point functions. In~\cite{Costa:2014rya}, they show that the number of three-point structures for general tensor operators is the same as the number of traceless-symmetric tensors (TSTs) of $SO(d)$ in
\beq
\bigotimes_{i=1}^3 \rho_i.
\eeq
This is equivalent to (\ref{eq:socounting}) because the only $SO(d)$ representations that give singlets after restriction to $SO(d-1)$ are TSTs, and each TST gives exactly one singlet.

We can also count the dimension of the conformal moduli space $\overline\MM_n=\MM_n/SO(d+1,1)$ of $n$ points, where $\MM_n$ is the configuration space of $n$ points on the sphere. By counting the unconstrained coordinates of the operators in conformal frame we get,
\beq
\dim\overline\MM_n=\sum_{i=2}^{m-2}\dim\alpha_i+d(n-m)=\frac{m(m-3)}{2}+d(n-m).
\eeq
This is of course also equal to
\beq
	\dim\overline\MM_n=\dim\MM_n-\dim SO(d+1,1)+\dim SO(d+2-m).
\eeq

\begin{examples}
Let us work out some simple examples of (\ref{eq:socounting}) in 3d. Let $\Bell$ denote the spin-$\ell$ representation of $SO(d)$, and $(s)$ denote the charge-$s$ representation of $SO(2)=U(1)$. For the trivial representation of the trivial group, we write $\bullet$.

Consider an $n$-point function of non-identical vectors in 3d. When $n=3$, the structures are given by $SO(2)$-singlets in
\begin{align}
\p{\Res^{SO(3)}_{SO(2)} \mathbf{1}}^{\otimes 3} &= \big((1) \oplus (0) \oplus (-1)\big)^{\otimes 3}\nn\\
 &= (3) \oplus 3(2)\oplus 6(1) \oplus 7(0) \oplus 6(-1) \oplus 3(-2) \oplus (-3).
\end{align}
In particular, there are 7 structures.

Let us emphasize that, despite the title of this paper, (\ref{eq:socounting}) actually gives the {\it space\/} of structures, not just the number. For example, consider a three-point function of vectors $J_i(\pol_i,x_i)=\pol_{i}^\mu J_{i\mu}(x_i)$, where $\pol_i^\mu$ are polarization vectors. Restricting to the conformal frame configuration $\<J_1(\pol_1,0)J_2(\pol_2,e_1)J_3(\pol_3,\oo)\>$, we can write seven invariants under the $SO(2)$ of rotations in the $2$-$3$ plane:
\begin{align}
\pol_{1}^1\pol_{2}^1\pol_{3}^1, \quad \pol_{1}^1\, \delta_{ab}\pol_{2}^a \pol_3^b,\quad \pol_{2}^1\, \delta_{ab} \pol_3^a \pol_1^b,\quad \pol_3^1\, \delta_{ab} \pol_1^a \pol_2^b,\nn\\
\pol_{1}^1\, \epsilon_{ab}\pol_{2}^a \pol_3^b,\quad \pol_{2}^1\, \epsilon_{ab} \pol_3^a \pol_1^b,\quad \pol_3^1\, \epsilon_{ab} \pol_1^a \pol_2^b,
\label{eq:explicitstructures}
\end{align}
where $\delta_{ab}$ and $\epsilon_{ab}$ are the two-dimensional metric and epsilon symbol.

The correlator is then given by (\ref{eq:cfreconstruction}). Alternatively, we can map the structures (\ref{eq:explicitstructures}) to the embedding-space structures of~\cite{Costa:2011mg} using the dictionary\footnote{Here, we use the nonstandard definition of an operator at infinity described in footnote~\ref{foot:noteaboutinfinity}.}
\begin{align}
\pol_i^1 &\mapsto V_{i}, \nn\\
\delta_{ab} \pol_i^a \pol_j^b &\mapsto H_{ij} + V_i V_j, \nn\\
\epsilon_{ab} \pol_i^a \pol_j^b &\mapsto 2\epsilon_{ij}.
\end{align}
The resulting expressions will automatically be free of redundancies.

 When $n\geq 4$, the stabilizer $SO(5-m)$ is trivial, and
\begin{align}
\p{\Res^{SO(3)}_{1} \mathbf{1}}^{\otimes n} &= \p{3\bullet}^{\otimes n} = 3^n \bullet,
\label{eq:vecexample3dgeneral}
\end{align}
so we have $3^n$ structures. In embedding space structures for $n\geq 5$, this corresponds to the fact that there are $3$ linearly-independent $V$ structures for each operator, and all $H$ structures are redundant. For $n=4$, we have two $V$ structures per point and the $H$ structures are replaced by $\epsilon(Z_i,P_1,P_2,P_3,P_4)$ in the notation of~\cite{Costa:2011mg}.

\end{examples}

\subsection{Parity}
\label{sec:parity}
If one wishes to distinguish parity-even and parity-odd structures, one has to note that the stabilizer group is actually $O(d+2-m)$ (for $n\geq 3$). There are two cases now, $n<d+2$ and $n\geq d+2$.

In the former case, $n<d+2$, the stabilizer subgroup contains a parity transformation. Therefore, parity of the correlator can be naturally defined on the conformal frame --- parity-even structures are scalars under $O(d+2-m)$ and parity-odd structures are pseudo-scalars. Another way to state this is that reflection fixes the conformal frame and thus all the conformal invariants $\mathbf{u}$ of $n$ points are parity even, and parity is a property of the tensor structure.

In the latter case, $n\geq d+2$, the stabilizer subgroup is trivial. Looking at the construction of the conformal frame, we see that parity actually acts within the conformal frame.\footnote{This is consistent with our definition of conformal frame, since that definition used only the connected component of the conformal group.} This means that there exist parity-odd conformal invariants $\mathbf{u}$ of $n$ points, and it is actually quite easy to construct one. In the embedding-space formalism of~\cite{Costa:2011mg} it can be written as
\beq
	\frac{\epsilon(P_1\cdots P_{d+2})}{\sqrt{P_{12}P_{23}\cdots P_{d+1,d+2}P_{d+2,1}}}.
\eeq
Note that the condition $n\geq d+2$ enters this construction naturally. Using this invariant, all the tensor structures can be chosen to be parity-even. Parity of the correlator is then the property of the coefficient functions $g^I$.\footnote{If in the definition of conformal frame we used the full conformal group, then parity would not act on the conformal frame, but it also would not be a part of the stabilizer. Rather, $r_x$ would contain the parity transformation for some $x_i$, and in that case the parity of the correlator would be supplied as extra information in the definition~\eqref{eq:cfreconstruction}.
}

\begin{examples}
Let us apply the above discussion to $n$-point functions of parity-even vectors in 3d. We denote the parity-even/odd spin-$\ell$ representations of $O(3)$ by $\Bell^\pm$. The spin-$\ell$ representations of $O(2)$ are denoted $\Bell$ and the scalars/pseudoscalars are denoted $\mathbf{0}^\pm$.\footnote{Though we sometimes use the same notation for representations of different groups (for example scalars/pseudoscalars of $O(2)$ and $O(3)$), we hope that the relevant group will be clear from context.}${}^,$\footnote{Note that spin-$\ell$ representations of $O(2)$ do not come in distinct parity-even and parity-odd versions. This is because $\epsilon_{\mu\nu}$ gives an isomorphism between the parity-even vector and the parity-odd vector in 2d. For spin-$\ell$ representations, we can act with $\epsilon_{\mu\nu}$ on one of the vector indices to get a parity-changing isomorphism. The only exception is the scalar representation, which comes in two versions $\mathbf{0}^\pm$, differing by a sign under reflections. Because of the $\epsilon$ isomorphism, we have $\mathbf{0}^\pm \otimes \Bell=\Bell$.} Finally, the parity-even/odd representations of $O(1)$ are denoted $\bullet^\pm$. For three-point functions, we have
\begin{align}
\p{\Res^{O(3)}_{O(2)} \mathbf{1}^+}^{\otimes 3} &= \p{\mathbf{1}\oplus \mathbf{0}^+}^{\otimes 3} = \mathbf{3} \oplus 3\,\mathbf{2} \oplus 6\, \mathbf{1} \oplus 4\, \mathbf{0}^{+} \oplus 3\, \mathbf{0}^{-},
\end{align}
so 4 of the 7 structures are parity-even and 3 are parity-odd, which is consistent with the explicit expressions~(\ref{eq:explicitstructures}). For four-point functions, we have
\begin{align}
\p{\Res^{O(3)}_{O(1)} \mathbf{1}^+}^{\otimes 4} &= \p{2\bullet^+ \oplus\, \bullet^-}^{\otimes 4}=41\,{\bullet^{+}} \oplus 40 \,{\bullet^{-}},
\end{align}
so $41$ of the $81$ structures are parity-even, and $40$ are parity-odd. For $n\geq 5$, parity-odd cross-ratios exist and all structures can be chosen to be parity even. This is easily seen to be in accordance with the discussion after \eqref{eq:vecexample3dgeneral}.
\end{examples}

\subsection{Permutation symmetry}
\label{sec:permutationsymmetry}
In this section we consider the constraints of permutation symmetries from the point of view of the conformal frame. Derivations of some technical results of this section are collected in appendix~\ref{app:permutations}.

Correlators involving identical operators are (anti-)symmetric under permutations of those operators.\footnote{In principle it might be interesting to consider also permutations which exchange non-identical operators, in order to switch between conformal frames differing only by the ordering of operators.} We can define the action of permutations on the correlator $g$ by
\beq
	(\pi g)^{a_1\ldots a_n}(x_1,\ldots,x_n)=\pm g^{a_{\pi(1)}\ldots a_{\pi(n)}}(x_{\pi(1)},\ldots, x_{\pi(n)}),
	\label{eq:permactiondef}
\eeq
with a $-$ sign for an odd permutation of fermions. In terms of polarizations,
\beq
	(\pi g)(\pol_i,x_i)=\pm g(\pol_{\pi(i)},x_{\pi(i)}).
\eeq
Invariance under a permutation $\pi$ is simply the statement that 
\beq
	\pi g=g.
	\label{eq:basicperminvariance}
\eeq
Of course, in order to impose this consistently with conformal invariance, the quantum numbers of the exchanged operators should be equal.

Applying a permutation $\pi$ to a conformal-frame configuration $p=\{x_i\}$ yields a new configuration $\pi p$ which is generically not in the conformal frame. To compare the value of the correlator at $\pi p$ with the value at $p$, one must find a conformal transformation that brings $\pi p$ back to the conformal frame. More precisely, choose for every $\pi$ a conformal transformation $r_\pi$ such that the configuration $x'_i=r_\pi^{-1}x_{\pi(i)}$ belongs to conformal frame (in general $r_\pi$ can depend on $x_i$). Then invariance \eqref{eq:confinv} and \eqref{eq:basicperminvariance} of the correlator requires
\beq
	r_\pi\pi g=g.
\eeq
By construction both the left hand side and right hand side depend only the values of $g$ on the conformal frame and thus this requirement can be phrased in terms of $g_0$.

Depending on whether $x'_i=x_i$, this either restricts the number of tensor structures allowed for $g_0$ by constraining its value at a single point of the conformal frame, or simply relates values of $g_0$ at different points in the conformal frame. An example of the latter case is the crossing-symmetry equation for four-point functions. In the former case we say that the permutation is ``kinematic''. The permutations which satisfy $x'_i=x_i$ (and thus preserve the cross-ratios $\mathbf{u}$) form a subgroup $S_n^\text{kin}\subseteq S_n$.

For $n\leq 3$ the conformal frame consists of a single point, so permutations simply give linear relations between tensor structures and we have $S_n^\text{kin}=S_n$. For four-point functions, $S_4^\text{kin}$ is the group of permutations that preserve $u$ and $v$. This is $S_4^\text{kin}=\mathbb{Z}_2^2=\{e,(12)(34),(13)(24),(14)(23)\}$ in cycle notation. For higher-point functions, $S^\text{kin}_n$ is trivial because no nontrivial permutation preserves all the cross-ratios.

Let us be more explicit and assume that the correlator is invariant under a subgroup $\Pi\subseteq S_n$. In terms of polarizations we have for any $\pi\in \Pi$, using \eqref{eq:conformalgroupaction} and \eqref{eq:permactiondef},
\beq	
	(r_\pi\pi g)(\pol_i,x_i)=(\pi g)(\RR_{r_\pi}(x_i)^{-1}\pol_i,r_\pi^{-1}x_i)\prod_{i=1}^n\Omega^{-\Delta_i}_{r_\pi}(x_i)=g(\pol'_i,x'_i)\prod_{i=1}^n\Omega^{-\Delta_i}_{r_\pi}(x_i),
\eeq
where 
\beq
	\pol_i'=\RR_{r_\pi}(x_{\pi(i)})^{-1}\pol_{\pi(i)},
\eeq
and the scaling factor with $\Omega$'s is trivial if the scaling dimensions are invariant under $\pi$, which we assume.
Suppose that the permutation is kinematic, $\pi\in\Pi^\text{kin}$, then the invariance condition becomes
\beq	
	g_0(\pol_i,x_i)=g_0(\pol'_i,x_i),
\eeq
and basically constrains the value of $g_0(\cdot, x_i)\in \bigotimes_i \rho_i$. Therefore, we see that there is an action of $\Pi^\text{kin}$ on $\bigotimes_i \rho_i$ which both permutes and twists the tensor factors. The tensor structures should be invariants of this action. 

Since only $S_3^\text{kin}$ and $S_4^\text{kin}$ are non-trivial, it is easy to consider the permutations on a case by case basis. We do this in appendix~\ref{app:permutations}. In particular we describe there all $r_\pi$ and the induced $\RR_{r_\pi}$, which are required for practical calculations with tensor structures. For example, we use these results in our account of 3d tensor structures in section~\ref{sec:3d}.

In the remainder of this section we derive group-theoretic rules for counting the permutation-symmetric tensor structures.

\subsubsection{Three-point structures}

In the case of three-point structures with non-trivial permutation symmetry we can have either $\Pi^\text{kin}=S_2$ or $\Pi^\text{kin}=S_3$. 

Let us start with $\Pi^\text{kin}=S_2$, where we have two identical operators $\OO_1=\OO_2$. Instead of going to the usual conformal frame, it is convenient to choose the configuration $\<\cO_1(-e) \cO_3(0) \cO_1(e)\>$, where $e$ is a unit vector. This gives a function $\tilde g(\pol_i,e)$. By analogy with the usual conformal frame, it is sufficient to ensure that $\tilde g(\pol_i,e)$ is covariant under $SO(d)$ rotations (where we allow $e$ to rotate as well as the $\pol_i$). 

Before taking permutation symmetry into account, the tensor structures are in one-to-one correspondence with traceless symmetric tensors in $\rho_1\otimes\rho_2\otimes\rho_3$. (As we explained in section~\ref{sec:npointconformalframe}, this is equivalent to the space of singlets in~\ref{eq:socounting}.) Each such tensor of spin $\ell$ can be contracted with $e_{\mu_1}\ldots e_{\mu_\ell}$ to give the corresponding $\tilde g$. Now, permutation symmetry demands
\beq
	\tilde g(\pol_1,\pol_2,\pol_3,e)=\pm \tilde g(\pol_2,\pol_1,\pol_3,-e)=\pm (-1)^\ell \tilde g(\pol_2,\pol_1,\pol_3,e),\label{eq:3ptS2perm}
\eeq
where the $\pm$ sign is determined by the statistics of the operators $\OO_1=\OO_2$, and the last equality is valid if $\tilde g$ comes from a spin-$\ell$ traceless-symmetric tensor in $\rho_1\otimes\rho_2\otimes\rho_3$. We find

\begin{proposition}[$S_2$]
\label{prop:stwo}
$S_2$-symmetric tensor structures are in one-to-one correspondence with even-spin traceless symmetric tensors in $\widehat{\mathrm{S}}^2\rho_1 \otimes \rho_3$ plus odd-spin traceless-symmetric tensors in $\widehat{\wedge}^2\rho_1 \otimes \rho_3$. Here, $\widehat{\mathrm{S}}^2$ denotes the symmetric square for bosonic arguments and exterior square for fermionic arguments, and $\widehat{\wedge}^2$ is defined analogously.
\end{proposition}

Now consider the case of $S_3$ symmetry with 3 identical operators. The full symmetry group is generated by permutations $(12)$ and $(123)$. We have already discussed $(12)$. We can generate the cyclic permutation $(123)$ by exponentiating the action of $(P^\mu + K^\mu) e_\mu$. This moves the operators along the line spanned by $e$ but does not rotate their polarizations, giving the condition
\beq
	\tilde g(\pol_1,\pol_2,\pol_3,e)=\tilde g(\pol_3,\pol_1,\pol_2,e).\label{eq:3ptS3perm}
\eeq 
Together,~\eqref{eq:3ptS2perm} and~\eqref{eq:3ptS3perm} give the trivial representation of $S_3$ when $\ell$ is even and the sign representation when $\ell$ is odd. This leads to 

\begin{proposition}[$S_3$]
\label{prop:sthree}
$S_3$-symmetric tensor structures are in one-to-one correspondence with even-spin traceless symmetric tensors in ${\mathrm{S}}^3\rho_1$ plus odd-spin traceless-symmetric tensors in ${\wedge}^3\rho_1$.\footnote{The distinction between $\hat S$ and $S$ has disappeared because all three operators are necessarily bosonic.}
\end{proposition}

In both propositions~\ref{prop:stwo} and~\ref{prop:sthree}, the parity of the structure is determined by the intrinsic parity of the traceless symmetric representations.

\subsubsection{Four-point structures}

Let us now count four-point structures. 
Recall that in the absence of permutation symmetries, the space of tensor structures is
\beq
\p{\mathrm{Res}^{O(d)}_{O(d-2)}\bigotimes_{i=1}^4 \rho_i}^{O(d-2)}.
\eeq
The most natural generalization to symmetric correlators would be to symmetrize the tensor product by the kinematic symmetries of the correlator, including factors of $(-1)$ for odd permutations of fermions. It turns out that this is almost correct, except that one does not need the $(-1)$'s. This is due to the fact that the conformal transformation that compares the permuted and unpermuted correlator also gives a $(-1)$ for an exchange of fermions. The general statement is 
\begin{proposition}[$\mathbb{Z}_2$ and $\mathbb{Z}_2^2$]
\label{prop:fourptperms}
The space of tensor structures for four-point functions with permutation symmetry $\Pi^\text{kin}$ is
\beq
	\p{\mathrm{Res}^{O(d)}_{O(d-2)}\p{\bigotimes_{i=1}^4 \rho_i}^{\Pi^\text{kin}}}^{O(d-2)},
\eeq
where $\Pi^\text{kin}$ acts by a simple permutation on the tensor factors, regardless of the fermion/boson nature of the operators, and the parentheses mean taking the invariant subspace.\footnote{One can also project to singlets of $\Pi^\text{kin}$ after applying $\Res^{O(d)}_{O(d-2)}$.}
\end{proposition}

We prove proposition~\ref{prop:fourptperms} in appendix~\ref{app:permutations4pt}.
There are two non-trivial options for $\Pi^\text{kin}$: $\mathbb{Z}_2$ and $\mathbb{Z}_2^2$. In the former case we simply need to compute the symmetric square of a representation. Indeed, without loss of generality assume that the non-trivial permutation is $(13)(24)$, and so $\rho_1=\rho_3$ and $\rho_2=\rho_4$. 
 It is easy to see that
\beq
\label{eq:z2prescription}
	\p{\bigotimes_{i=1}^4 \rho_i}^{\mathbb{Z}_2}=\mathrm{S}^2(\rho_1\otimes \rho_2)
\eeq
The latter case is a bit more involved. First, all the representations have to be identical, $\rho_1=\rho_2=\rho_3=\rho_4=\rho$. The relevant formula is then, as we show in appendix~\ref{app:characters},
\beq
\label{eq:z2sqprescription}
	\p{\bigotimes_{i=1}^4 \rho_i}^{\mathbb{Z}_2^2}=\rho^4\ominus 3\p{\wedge^2\!\rho\otimes\mathrm{S}^2\rho},
\eeq
where $\ominus$ represents the formal difference\footnote{\label{rem:characterdifference}One can think about representations in terms of characters. Since characters are functions, there is no problem with taking differences. Alternatively, one can think of a reducible representation as a formal sum of irreducible representations with non-negative coefficients. Then, taking a difference of representations is equivalent to taking differences of these coefficients. Some coefficients may end up being negative, in which case the result is called a ``virtual" representation. The representation~\eqref{eq:z2sqprescription} is guaranteed not to be virtual.} in the character ring. 

\begin{examples}
As examples, consider $n$-point correlators of identical parity-even vectors in 3d. For $n=3$, we have the following identities among $O(3)$ representations:
\begin{align}
\mathrm{S}^3 \mathbf{1}^+ &= \mathbf{3}^+,\nn\\
\wedge^3 \mathbf{1}^+ &= \mathbf{0}^-.
\end{align}
By proposition~\ref{prop:sthree}, it follows that there are no nontrivial three-point structures. For $n=4$, using proposition~\ref{prop:fourptperms} with $\Pi^\text{kin}=\mathbb{Z}_2^2$ and equation~(\ref{eq:z2sqprescription}), we have
\begin{align}
(2\,{\bullet^{+}}\oplus {\bullet^{-}})^4 \ominus 3\p{\wedge^2 (2\,{\bullet^{+}}\oplus {\bullet^{-}}) \otimes \mathrm{S}^2(2\,{\bullet^{+}}\oplus {\bullet^{-}})} &= 17\,{\bullet^+} \oplus 10\,{\bullet^-},
\end{align}
so there are 17 parity-even structures and 10 parity-odd structures in a four-point function of identical vectors. Finally, for $n\geq 5$, kinematic permutations are absent, so there are $3^n$ structures (which can be taken to be parity-even).

Consider an example with two identical Majorana fermions and two identical scalars, all parity-even. We have the following $O(2,1)$ identity
\beq
\mathrm{S}^2 \spinor = \mathbf{1}^+.
\eeq
Using proposition~\ref{prop:fourptperms} with $\Pi^\text{kin}=\mathbb{Z}_2$ and equation~(\ref{eq:z2prescription}), we find the space of four-point structures
\beq
	2\,{\bullet^{+}}\oplus {\bullet^{-}},
\eeq
so there are 2 parity-even structures and 1 parity-odd structure. This agrees with~\cite{Iliesiu:2015akf}. Note that it was essential not to include $(-1)$ for a permutation of fermions in proposition~\ref{prop:fourptperms}. 

\end{examples}

\subsection{Summary: tensor structures of long-multiplets}
\label{sec:generalcountingrule}

The discussion above can be summarized as the following theorem.
\begin{theorem}
\label{thm:main}
The conformal correlator involving $n\geq 3$ operators in representations $\rho_i$ can be written as
\beq
\crr{\OO_1^{a_1}(x_1)\ldots\OO_n^{a_n}(x_n)}=\sum_I \mathbb{Q}^{a_1\ldots a_n}_I g^I(\mathbf{u}),\label{eq:thmcorr}
\eeq
where $\mathbf u$ is a set of coordinates on the  conformal moduli space $\overline\MM_n$ of $n$ points $x_1\ldots x_n$,
\beq
	\dim\overline\MM_n=\frac{m(m-3)}{2}+d(n-m),\quad m = \min(n,d+2),
\eeq
and the conformally-invariant tensor structures $\mathbb{Q}_I$ are in one-to-one correspondence with scalars (for parity-even structures) and pseudo-scalars (for parity-odd structures) in the representation of $O(d+2-m)$ given by
\beq
	\mathrm{Res}_{O(d+2-m)}^{O(d)}\bigotimes_{i=1}^n\rho_i.\label{eq:thmrule}
\eeq
If parity is not conserved, one simply replaces $O(\cdot)$ groups with $SO(\cdot)$ groups above. If $n\geq d+2$, then one can form parity-odd cross-ratios, and parity of the correlator is rather a property of the functions $g^I$ rather that the structures $\mathbb{Q}_I$, which can all be chosen to be parity-even. 

When $n=3$ or $n=4$ the correlator~\eqref{eq:thmcorr} can have a group $\Pi^\text{kin}$ of permutation symmetries which leave $\mathbf{u}$ invariant, and thus impose constraints on the structures $\mathbb{Q}_I$. The spaces of structures in these cases are described in propositions~\ref{prop:stwo},~\ref{prop:sthree}, and~\ref{prop:fourptperms}.

\end{theorem}

\section{Conservation conditions}
\label{sec:conserved}
We now consider correlation functions of operators that satisfy conservation conditions. We are mainly interested in the number of ``functional degrees of freedom" in such correlators --- i.e.\ the number of functions of cross-ratios needed to completely specify the correlator~\cite{Dymarsky:2013wla}. For simplicity, we mostly restrict our attention to traceless symmetric tensor conserved currents, of which spin-$1$ currents and the stress tensor are prime examples. We describe the modifications required for more general operators at the end of this section. 

Correlation functions involving conserved currents are constrained by differential equations such as
\beq
	\frac{\partial}{\partial x_1^{\mu_1}}\crr{J^{\mu_1\ldots\mu_\ell}(x_1)\ldots}=\frac{\partial}{\partial x_1^{\mu_1}}\sum_{I=1}^N \mathbb{Q}^{\mu_1\ldots \mu_\ell\ldots}_I(x_i) g^I(\mathbf{u})=\text{ contact terms}.\label{eq:conservationsingle}
\eeq
When $n\geq 4$, these are differential constraints on the functions $g_I(\mathbf{u})$. In general, the full set of conservation equations is not independent and this makes it not immediately clear how many degrees of freedom there actually are. The purpose of this section is to classify the relations between these equations and motivate a group-theoretic rule for the number of degrees of freedom of such correlators for $n\geq 4$.

Our rule will also classify ``generic" three-point functions --- i.e.\ three-point correlators where at least one operator has generic dimension $\Delta$. When the dimensions of operators are non-generic, extra three-point structures can appear. The simplest example occurs for a three-point function of a conserved current and two scalars, $\< J_\mu \phi_1 \phi_2\>$. Generically, no structure exists for such a correlator, but a special structure becomes possible when the scalars have equal dimensions $\Delta_1=\Delta_2$. These special structures are related to the contact terms on the right-hand side of (\ref{eq:conservationsingle}). There is a beautiful story behind them that we explore in~\cite{KravchukFuture}. (In particular, we complete the classification of conformally-invariant tensor structures in~\cite{KravchukFuture}.) For higher-point correlators, non-generic structures have a fixed $x_i$ dependence, so they do not contribute to the number of functional degrees of freedom.

Our strategy is to understand the relations between equations~\eqref{eq:conservationsingle}. In general, if we have a system of equations
\beq
	D_1\mathbf{g}=0,\label{eq:Dgequations}
\eeq
where $\mathbf{g}$ is a vector of $N_0$ unknown functions and $D_1$ is a $N_1\times N_0$ matrix with differential operator coefficients, we say that there are relations between the equations~\eqref{eq:Dgequations} if there is an $N_2\times N_1$ matrix $D_2$ such that 
\beq
	D_2D_1=0.
\eeq
Note that here $D_2D_1\mathbf{g}=0$ independently of~\eqref{eq:Dgequations}. There is a sense in which $D_2$ can be complete. Namely, we say that $D_2$ is a compatibility\footnote{This name comes from considering the equation $D_1\mathbf{g}=\mathbf{f}$. The function $\mathbf{f}$ is compatible with this equation only if $D_2\mathbf{f}=0$. Systems of equations for which a non-trivial $D_2$ exists are known as overdetermined systems.} operator for $D_1$ iff any other $\tilde D_2$ satisfying $\tilde D_2 D_1=0$ can be expressed as $\tilde D_2=QD_2$ for some matrix differential operator $Q$.
It can happen that there are further relations between the relations $D_2$, i.e.\ an $N_3\times N_2$ matrix $D_3$ such that
\beq
	D_3D_2=0,\text{ etc.}
\eeq
If at some point this sequence of compatibility operators terminates --- i.e.\ for $i>i_0$ we have $N_i=0$ --- then we can compute a version of the Euler characteristic
\beq
	N=\sum_{i=0}^\infty (-1)^iN_i.
\eeq
We expect that $N$ is the true number of functional degrees of freedom parametrizing a solution to~\eqref{eq:Dgequations}. Note that by the number of functional degrees of freedom we mean the functional parameters which depend on the same number of variables as the original equation.

Consider first the simplest case of conservation of a spin-$\ell$ traceless-symmetric current,
\beq
\frac{\partial}{\partial x^{\mu_1}}J^{\mu_1\ldots\mu_\ell}(x)=0,\label{eq:conservationJ}
\eeq
which can be phrased as setting to zero a spin-$(\ell-1)$ operator
\beq
	V^{\mu_1\ldots\mu_{\ell-1}}(x)=\frac{\partial}{\partial x^{\mu}}J^{\mu\mu_1\ldots\mu_{\ell-1}}(x).
\eeq
If the current $J$ has scaling dimension $\Delta_J=d+\ell-2$, then the conservation equation is conformally-covariant, meaning simply that $V$ transforms as a primary operator. Note that $V$ is still conserved, but $\partial V=0$ does not constitute a relation between the conservation equations 
in the above sense
 --- it only holds if the original equation is satisfied. In fact, there is no differential operator which annihilates the left hand side of~\eqref{eq:conservationJ}.

Since $V$ is a primary, inserting it into a correlator we find 
\beq
	\crr{V^{\mu_1\ldots\mu_{\ell-1}}\ldots}=\sum_{I=1,J=1}^{N_1,N}\tilde{\mathbb{Q}}_I^{\mu_1\ldots\mu_{\ell-1}\ldots} (D_1)^I{}_J g^J(\mathbf{u})=0,
\label{eq:conservationV}
\eeq
where the structures $\tilde{\mathbb{Q}}_I$ are the conformally invariant structures suitable for the correlator on the left. Note that the structures $\mathbb{Q}$ are in one-to-one correspondence with singlets in 
\beq
	[\ell\otimes\rho_2\otimes\ldots]=[\ell]\otimes[\rho_2]\otimes\ldots,
\eeq
where we use $[\,\cdot\,]$ to denote the restriction to $SO(d+2-m)$. On the other hand, the structures $\tilde{\mathbb{Q}}$ are given by the singlets in
\beq
	[(\ell-1)\otimes\rho_2\otimes\ldots]=[\ell-1]\otimes[\rho_2]\otimes\ldots.
\eeq
If there is only one current in the correlator, then there are no relations between the equations and the number of degrees of freedom is given by the number of singlets in
\beq
	\Big([\ell]\otimes[\rho_2]\otimes\ldots\Big)\ominus\Big([\ell-1]\otimes[\rho_2]\otimes\ldots\Big)=\Big([\ell]\ominus[\ell-1]\Big)\otimes[\rho_2]\otimes\ldots.
\eeq
Here the $\ominus$ is the formal difference\footnote{See footnote~\ref{rem:characterdifference}.} in the character ring of $SO(d+2-m)$. The idea now is to note
\beq
	\mathrm{Res}^{SO(d)}_{SO(d-1)}\ell\ominus\mathrm{Res}^{SO(d)}_{SO(d-1)}(\ell-1)=\ell',
\eeq
where $\ell'$ is the spin-$\ell$ traceless symmetric representation of $SO(d-1)$.\footnote{Note that $SO(d-1)$ is the little group for massless particles in $d+1$ dimensions. We will make use of this fact in section~\ref{sec:scattering}.} Therefore, we see that the number of degrees of freedom is given by the singlets in
\beq
	[\ell']\otimes[\rho_2]\otimes\ldots\label{eq:conservedcounting}
\eeq
One may wonder if this rule holds more generally --- i.e.\  whether one can compute the number of degrees of freedom in any correlator involving conserved operators by simply replacing the $SO(d)$ representations of these operators with their ``effective'' $SO(d-1)$ representations in Theorem~\ref{thm:main}. This is indeed so\footnote{As we note in the beginning of this section, for three point functions this is only true for sufficiently generic scaling dimensions of the operators.}, and in section~\ref{sec:conservedmultisym} we show in examples how this rule works in the situations when we have several conserved operators or when there are permutation symmetries.

In the example considered above the primary $V$ obtained  from $J$ did not have any null states of its own, so it was easy to count the number of degrees of freedom in the correlator~\eqref{eq:conservationV}. For operators $J$ satisfying more general conformally-invariant differential equations it may turn out that $V$ itself has a null descendant $V'$, and thus satisfies a conformally-invariant differential equation expressed as $V'=0$. Now $V'$ can turn out to have null descendants $V''$, and so on. A simple class of examples when this happens are the differential forms from the de Rham complex. Repeating the above analysis, we see that the effective $SO(d-1)$ representation we should use in this situation is
\beq
	[\rho]\ominus[v]\oplus[v']\ominus[v'']\oplus \ldots,\label{eq:effectiverepresentation}
\eeq 
where $\rho$ is the $SO(d)$ representation of $J$ and $v$ is the $SO(d)$ representation of $V$ and so on.

We expect that quite generally this alternating sum gives an actual representation of $SO(d-1)$. Indeed, we have $V=\mathcal{D}J$ for some conformally invariant differential operator $\mathcal{D}$. Because of translation invariance $\mathcal{D}$ has constant coefficients, and thus the equation 
\beq
	\mathcal{D}J=0
\eeq
is in momentum space a simple linear equation for the amplitude $J$. In particular, for each fixed momentum $p$, the space of solutions is a finite-dimensional representation of $SO(d-1)$ which leaves $p$ invariant. It is easy to convince oneself that this is the representation which~\eqref{eq:effectiverepresentation} is computing.

In applications to unitary conformal field theories we are only interested in operators $J$ with the scaling dimension saturating some unitarity bound --- these are the only operators which are unitary and have null descendants at the same time. A detailed classification of such operators can be found in section 5 of~\cite{Cordova:2016emh} (see also~\cite{Minwalla:1997ka,Siegel:1988gd}), here we only give a short summary. Among these operators, some can be classified as free and the rest, which we will call the unitary conserved currents, satisfy first-order differential equations. In 3d and 4d all unitary conserved currents are generalizations of $(d-1)$-forms and they do not have the analogue of $V'$. In 5d and 6d there appear unitary conserved currents which generalize $(d-2)$-forms, and they have $V'$ but not $V''$. Given the classification in \cite{Cordova:2016emh}, it is an easy exercise to find the effective $SO(d-1)$ representation for arbitrary unitary conserved currents in $d\leq 6$.

\subsection{Multiple conserved operators and permutation symmetries}
\label{sec:conservedmultisym}
Let us see how the rule~\eqref{eq:conservedcounting} behaves when there are several conserved currents in the correlator. Consider for example the case of two currents $J_1$ and $J_2$. We then have the equations
\al{
	\crr{V_1J_2\ldots}&=0,\label{eq:eqJV1}\\
	\crr{J_1V_2\ldots}&=0.\label{eq:eqJV2}
}
But there is a relation between these equations. Taking the remaining divergences in both equations we arrive in both cases at 
\al{
	\crr{V_1V_2\ldots}&=0,\label{eq:eqVV}
}
and by taking the difference we obtain $0$ regardless of whether $V_i=0$ or not. This thus leads to a number of relations. This number is equal to the number of tensor structures in $\crr{V_1V_2\ldots}$. Therefore, we need to add it to the number of degrees of freedom,
\begin{align}
\Big([\ell_1]\otimes[\ell_2]\Big)\ominus\Big([\ell_1]\otimes[\ell_2-1]\Big)\ominus\Big([\ell_1-1]\otimes[\ell_2]\Big)\oplus\Big( [\ell_1-1]\otimes[\ell_2-1]\Big)=[\ell_1']\otimes[\ell_2'].
\end{align}
It is easy to see that this generalizes to any number of conserved operators. 

Consider now the case when the operators $J_1$ and $J_2$ are identical, $\ell_1=\ell_2=\ell$ and there is a kinematic permutation expressing this. Assume that $n=4$ and the other operators are scalars for simplicity. In this case the equations~\eqref{eq:eqJV1} and~\eqref{eq:eqJV2} are equivalent, since the tensor structures for $\crr{J_1J_2\ldots}$ are chosen to be symmetric. Then we can use just one equation, say~\eqref{eq:eqJV1}. However, it is still subject to relations. In particular, if we take an extra divergence to get to the equation~\eqref{eq:eqVV}, we will find that it is symmetric in permutation of $V$'s, and thus antisymmetrizing the $V$'s we get $0$. Since it is a non-trivial operation which we applied to~\eqref{eq:eqJV1}, it constitutes a relation among equations~\eqref{eq:eqJV1}. Therefore we need to look for scalars in 
\beq
	\mathrm{S}^2[\ell]\ominus\Big([\ell]\otimes[\ell-1]\Big)\oplus \wedge^2[\ell-1].
\eeq
Incidentally, the following relation holds in the character ring,
\beq
	\mathrm{S}^2\left(\chi_1-\chi_2\right)=\mathrm{S}^2\chi_1-\chi_1\chi_2+\wedge^2 \chi_2.
\eeq
It can be easily derived from the character formulas~\eqref{eq:symmetriccharacter} and~\eqref{eq:wedgecharacter}. We therefore see that the prescription works even when there is a permutation symmetry,
\beq
	\mathrm{S}^2[\ell']=\mathrm{S}^2[\ell]\ominus\Big([\ell]\otimes[\ell-1]\Big)\oplus \wedge^2[\ell-1].
\eeq
The techniques above also allow us to keep track of parity by simply replacing $SO$ groups with $O$ groups.

\begin{examples}[Conserved four-point functions in 3d and 4d]
As examples, let us compute the number of functional degrees of freedom in a four-point function of identical, conserved, parity-even, spin-$\ell$ currents in 3d and 4d.  Applying proposition~\ref{prop:fourptperms}, equation (\ref{eq:z2sqprescription}), and the discussion above, we must find the number of $O(d-2)$ scalars $\bullet^+$ and pseudoscalars $\bullet^-$ in
\beq
\label{eq:identicalcurrentrepresentation}
\rho = [\ell']^4 \ominus 3\p{\wedge^2[\ell'] \otimes \mathrm{S}^2 [\ell']}.
\eeq

In 3d, $[\ell']$ is the restriction of the spin-$\ell$ traceless symmetric tensor of $O(2)$ to $O(1)$, which is simply $[\ell']={\bullet^+}\oplus{\bullet^-}$. Plugging in we easily find
\beq
\rho_{\textrm{3d}} = 5\,{\bullet^+} \oplus 2\,{\bullet^-},
\eeq
so there are 5 parity even and 2 parity odd degrees of freedom. Note that the answer is independent of $\ell$. As we will see in section~\ref{sec:scattering}, this is related to the fact that massless particles in 4d always have two degrees of freedom, regardless of helicity.

In 4d, it is convenient to use characters of $O(2)$. $O(2)$ is a semidirect product 
\beq
U(1)\rtimes \mathbb{Z}_2 = \{(x,s): x\in U(1), s=\pm 1 \},
\eeq
with the multiplication rule
\beq
(x_1,s_1)(x_2,s_2) = (x_1 x_2^{s_1}, s_1 s_2).
\eeq
The spin-$j$ representation $\mathbf{j}$ has character
\beq
\label{eq:o2spinellchar}
\chi_{\Bell}(x,s) = \frac{1+s}{2}(x^j + x^{-j}),
\eeq
while the scalars $\bullet^+$ and pseudoscalars $\bullet^-$ have characters $1$ and $s$, respectively.  $[\ell']$ is the restriction of the parity-even spin-$\ell$ representation of $O(3)$ to $O(2)$, namely
\beq
[\ell'] = \Bell \oplus (\Bell-\mathbf{1}) \oplus \dots \oplus \mathbf{1} \oplus \bullet^+,
\eeq
which has character
\beq
\label{eq:dimreductionchar}
\chi_{[\ell']}(x,s) = \frac{1+s}{2}\frac{x^{\ell+\frac 1 2}- x^{-\ell-\frac 1 2}}{x^{\frac 1 2} - x^{-\frac 1 2}} + \frac{1-s}{2}.
\eeq
Plugging (\ref{eq:dimreductionchar}) into equation~(\ref{eq:characterz2z2}) for the character of a $\mathbb{Z}_2^2$-invariant tensor product, we find
\begin{align}
\chi_{\rho_\textrm{4d}}(x,s) &= \frac{1+s}{2} \left(\frac 1 4\left(\frac{x^{\ell+\frac 1 2}-x^{-\ell-\frac 1 2}}{x^{\frac 1 2}-x^{-\frac 1 2}}\right)^4+\frac 3 4\left(\frac{x^{2 \ell+1}-x^{-2\ell-1}}{x-x^{-1}}\right)^2\right)+ \frac{1-s}{2}(3 \ell^2 + 3\ell+1) \nn\\
&= \frac{(4\ell +3)(\ell+2)(\ell+1)}{6} + \frac{(4\ell+1)\ell(\ell-1)}{6}s + \dots,
\label{eq:characterforidenticalspinlinfourd}
\end{align}
where ``$\dots$" represents sums of spin-$j$ characters (\ref{eq:o2spinellchar}). The constant term in (\ref{eq:characterforidenticalspinlinfourd}) is the number of parity-even structures and the coefficient of $s$ is the number of parity-odd structures. Plugging in $\ell=1,2$, we obtain $7+0s$ and $22+3s$, respectively, in agreement with~\cite{Dymarsky:2013wla}.

\end{examples}

\section{Correlation functions in 3d}
\label{sec:3d}
In this section we consider in detail correlation functions in three dimensions, in order to exemplify how our formalism gives the tensor structures rather than just their number, and how this can be applied in practice.

\subsection{Conventions for $SO(2,1)$}
\label{sec:3dintro}
In this section we will be working in Lorentzian signature in order to allow Majorana spinors. Our conventions for spinors will be those of~\cite{Iliesiu:2015qra}. In this subsection we describe the basic notation. 

The primary operators in $2+1$ dimensions transform in representations of $Spin(2,1)\simeq Sp(2,\mathbb R)$. The smallest such representation is the two-component Majorana spinor $\mathbf{\frac{1}{2}}$, the fundamental of $Sp(2,\mathbb R)$
\beq
	\psi^\alpha.
\eeq
This representation is equivalent to its dual 
\beq
	\psi_\alpha,
\eeq
due to the invariant symplectic form of $Sp(2,\mathbb R)$
\beq
	\Omega^{\alpha\beta}=\Omega_{\alpha\beta}=\begin{pmatrix}
	0 & 1 \\ -1 & 0
	\end{pmatrix},\quad \psi_\alpha=\Omega_{\alpha\beta}\psi^\beta.
\eeq
We have $\spinor\otimes\spinor=\mathrm{S}^2\spinor\oplus\wedge^2\spinor=\mathbf{1}\oplus\mathbf{0}$. The equivalence between $S^2\spinor$ and the vector representation of $Spin(2,1)$ is established by the gamma matrices $(\gamma^\mu)^\alpha{}_\beta$,
\beq
	\gamma^0=\begin{pmatrix}
	0 & 1 \\ -1 & 0
	\end{pmatrix},\quad
	\gamma^1=\begin{pmatrix}
	0 & 1 \\ 1 & 0
	\end{pmatrix},\quad
	\gamma^2=\begin{pmatrix}
	1 & 0 \\ 0 & -1
	\end{pmatrix}.
\eeq
More precisely, we have
\beq
	v^\mu=\Omega_{\alpha\sigma}(\gamma^\mu)^\sigma{}_\beta v^{(\alpha\beta)}.\label{eq:vectorspinor}
\eeq
Generally, all finite-dimensional representations of $Spin(2,1)$ are the symmetric powers of the Majorana representation, $\Bell=\mathrm{S}^{2\ell}\spinor$. We therefore represent an arbitrary real operator $\OO$ of spin $\ell$ as
\beq
\OO^{(\alpha_1\ldots \alpha_{2\ell})}(x),	
\eeq
and we will use index-free notation by introducing a polarization spinor $s$,
\beq
	\OO(s,x)=s_{\alpha_1}\ldots s_{\alpha_{2\ell}}\OO^{(\alpha_1\ldots \alpha_{2\ell})}(x).
\eeq

We need to make a choice of $Pin(2,1)$ group to consider parity. Reflection $x^1\to - x^1$ is generated by
\beq
	\psi\to \pm\gamma^1\psi,
\eeq
and reflection $x^2\to -x^2$ is generated by
\beq
	\psi\to \pm\gamma^2\psi,
	\label{eq:3dparity2}
\eeq
as can be checked by considering the induced action on the vector representation. The sign ambiguity reflects the fact that it is a double cover $Pin(2,1)$ of $O(2,1)$ which acts on spinors, so there are twice as many ``reflections'' as in $O(2,1)$.

\subsection{Three-point structures}
\label{sec:3d3pt}
We choose the standard positions for the three operators by picking 
\al{
	x_1&=(0,0,0),\\
	x_2&=(0,0,1),\\
	x_3&=(0,0,L),
}
and considering the correlator
\beq
	g_0(s_1,s_2,s_3)=\lim_{L\to+\infty}L^{2\Delta_3}\crr{\OO_1(s_1,x_1)\OO_2(s_2,x_2)\OO_3(s_3,x_3)}.
\eeq

The connected component of the stabilizer subgroup in this case consists of boosts $s_i\to e^{-i \lambda \overline K_1}s_i$ with
\beq
	\overline K_1=\frac{1}{2}
	\begin{pmatrix}
		i & 0 \\ 0 & -i	
	\end{pmatrix}.
\eeq
Writing 
\beq
	(s_i)_\alpha=\begin{pmatrix}
	\xi_i \\ \bar\xi_i,
	\end{pmatrix}\label{eq:xi}
\eeq
we see that $\xi_i$ has charge $+1/2$ under these boosts, and $\bar\xi_i$ has charge $-1/2$. 

According to the general rule, the three-point functions are in one-to-one correspondence with stabilizer-invariant functions $g_0(s_i)$. Clearly, one can choose a basis for such functions consisting of monomials 
\beq
	[q_1q_2q_3]=\prod_{i=1}^3 \xi_i^{\ell_i+q_i}\bar \xi_i^{\ell_i-q_i},
\eeq
with $q_i\in\{-\ell_i,\ldots,\ell_i\}$ subject to
\beq
	\sum_{i=1}^3 q_i=0
	\label{eq:3dthreepointneutrality}
\eeq

If parity is conserved, then stabilizer subgroup also contains parity transformation $s_i\to \gamma_1 s_i$. This simply exchanges $\xi_i$ and $\bar\xi_i$. Therefore, structures of definite parity are given by
\beq
	[q_1q_2q_3]^\pm \equiv \prod_{i=1}^3 \xi_i^{\ell_i+q_i}\bar \xi_i^{\ell_i-q_i}\pm\prod_{i=1}^3 \xi_i^{\ell_i-q_i}\bar \xi_i^{\ell_i+q_i},
	\label{eq:3dthreepointparity}
\eeq
and now sets $q_i$ and $-q_i$ are identified.

\paragraph{Permutations.} Consider the permutations, starting with the transposition $(12)$. According to the general rule, we need to apply a transformation which brings the operators back to the conformal frame position after the permutation. We are interested in the $Spin(3)$ elements
\beq
	\RR_{r_\pi}^{-1}(x_i)
\eeq
induced at the insertions of the operators. These are computed in the appendix~\ref{app:permutations} with the result that for all transpositions there are $e^{\pm i\pi/2}$ at all insertions, inducing $s_i\mapsto \pm \gamma^0 s_{\pi(i)}$, under which $\xi_i\mapsto\pm\bar\xi_{\pi(i)}$ and $\bar\xi_i\mapsto\mp\xi_{\pi(i)}$. Taking into account the precise signs, we find the action of the permutations
\al{
	(12):\,&[q_1q_2q_3]^\pm\mapsto \pm(-1)^{\ell_1+\ell_2-\ell_3}[q_2q_1q_3]^\pm,\label{eq:3perm12}\\
	(13):\,&[q_1q_2q_3]^\pm\mapsto \pm(-1)^{\ell_1+\ell_2+\ell_3}[q_3q_2q_1]^\pm,\label{eq:3perm13}\\
	(23):\,&[q_1q_2q_3]^\pm\mapsto \pm(-1)^{-\ell_1+\ell_2+\ell_3}[q_1q_3q_2]^\pm\label{eq:3perm23}.
}
If the permutations are symmetries of the correlator, the signs in front of $\ell_i$ above can all be chosen to be $+$, since e.g.\ for permutation $(12)$ $\ell_3$ has to be integral for the full correlator to be bosonic. Under these permutations the tensor structure has to be symmetric or anti-symmetric depending on whether the exchanged operators are bosons or fermions. Redefining the permutations as
\al{
	(12)':\,&[q_1q_2q_3]^\pm\mapsto \pm(-1)^{\ell_3}[q_2q_1q_3]^\pm,\label{eq:3perm12f}\\
	(13)':\,&[q_1q_2q_3]^\pm\mapsto \pm(-1)^{\ell_2}[q_3q_2q_1]^\pm,\label{eq:3perm13f}\\
	(23)':\,&[q_1q_2q_3]^\pm\mapsto \pm(-1)^{\ell_1}[q_1q_3q_2]^\pm\label{eq:3perm23f},
}
we now have the requirement that the tensor structure is symmetric regardless of the nature of the operators.

\paragraph{Counting.} Let us now count the number of structures, assuming all the operators to be different. By counting all possible combinations of $q_i$ one easily recovers the result of~\cite{Costa:2011mg} for the number of 3-point structures, 
\beq
	N_{3d}(\ell_1,\ell_2,\ell_3)=(2\ell_1+1)(2\ell_2+1)-p(p+1),
\eeq
where $p=\max(\ell_1+\ell_2-\ell_3,0)$ and $\ell_1\leq \ell_2\leq \ell_3$. Unless all three operators are bosons, $q_i\equiv 0$ is not a solution, and thus there is an equal number of parity-even and parity-odd structures. In case all three operators are bosons, $q_i\equiv 0$ gives a valid parity-even structure. In this case the number of parity-even structures is larger than the number of parity-odd structures by $1$. We then have for the number of definite-parity structures
\beq
	N_{3d}^\pm(\ell_1,\ell_2,\ell_3)=\frac{N_{3d}(\ell_1,\ell_2,\ell_3)\pm\kappa}{2}\label{eq:3ptnparity},
\eeq
where $\kappa=1$ when all the operators are bosonic, and $\kappa=0$ otherwise.

In the case when there are identical operators, there are two options. The first option is that there are two identical operators, say $\ell_1=\ell_2$. The second is that all three operators are identical. In the first case one can show
\al{
	N_{3d}^\pm(\ell_1\leftrightarrow\ell_2,\ell_3)=&\frac{N^\pm_{3d}(\ell_1,\ell_1,\ell_3)}{2}+\frac{(-1)^{\ell_3}}{2}\left[\ell_1+\tfrac{1\pm\kappa}{2}\pm\min(\floor{\ell_1+\tfrac{1}{2}},\floor{\tfrac{\ell_3+1-\kappa}{2}})\right],\label{eq:12symcount}
}
and in the second case
\beq
	N^\pm_{3d}(\ell)=\frac{1}{6}\left[N^\pm_{3d}(\ell,\ell,\ell)+(-1)^\ell\left(3\ell+\tfrac{3}{2}\pm 3\floor{\tfrac{\ell}{2}}\pm\tfrac{3}{2}\right)+1\pm 1\right].\label{eq:123symcount}
\eeq
These formulas can be obtained either from propositions~\ref{prop:stwo},~\ref{prop:sthree} and character formulas of appendix~\ref{app:characters} or from the above description of permutations by computing the character of $S_2$ or $S_3$ on the space of tensor structures $[q_1q_2q_3]^\pm$.

\subsection{Four-point structures}
\label{sec:3d4pt}
For four operators, we choose the following conformal frame
\al{
	x_1&=(0,0,0),\\
	x_2&=(t,x,0),\\
	x_3&=(0,1,0),\\
	x_4&=(0,L,0),
}
and consider the correlator
\beq
	g_0(s_i,t,x)=\lim_{L\to+\infty}L^{2\Delta_4}\crr{\OO_1(s_1,x_1)\OO_2(s_2,x_2)\OO_3(s_3,x_3)\OO_3(s_4,x_4)}.
\eeq
We will mostly use the parameters
\beq
	z=x-t,\quad \bar z = x+t,
\eeq
such that under the continuation to Euclidean time $t_E=it$, we will get the usual holomorphic and anti-holomorphic coordinates.

Note that the stabilizer subgroup is just the $O(1)$ of reflections $x^2\to -x^2$. Therefore, any function of $s_i$ with appropriate homogeneous degrees will give us a valid 4-point structure. More precisely, we can write
\beq
g_0(s_i,z,\bar z) = \sum_{q_i}[q_1q_2q_3q_4] g_{[q_1q_2q_3q_4]}(z,\bar z),
\eeq
where 
\beq
	[q_1q_2q_3q_4]=\prod_{i=1}^4 \xi_i^{\ell_i+q_i}\bar\xi_i^{\ell_i-q_i}
\eeq
with $\xi$, $\bar \xi$ as in~\eqref{eq:xi} and $q_i\in\{-\ell_i\ldots\ell_i\}$.

The action of spatial parity is, according to \eqref{eq:3dparity2}, $s_i\mapsto\gamma_2 s_i$ or $\xi_i\mapsto \xi_i$, $\bar\xi_i\mapsto-\bar\xi_i$. Therefore,
\beq\label{eq:parity3dfourpoint}
	[q_1q_2q_3q_4]\mapsto(-1)^{\sum_i \ell_i-q_i}[q_1q_2q_3q_4].
\eeq
We see that the structures we have chosen already have definite parity.

\paragraph{Permutations and crossing symmetry.} Consider now how the four-point functions transform under the permutations. Since we are working in Lorentzian signature now, we need to perform an analytic continuation of the phases in appendix~\ref{app:permutations}.
Doing this, we obtain the following formulas for the nontrivial permutations,
\al{
	(12)(34)\,&:\,[q_1q_2q_3q_4]\mapsto n((z-1)^{q_1+q_4-q_2-q_3})[q_2q_1q_4q_3],\\
	(13)(24)\,&:\,[q_1q_2q_3q_4]\mapsto n(z^{q_3+q_4-q_1-q_2}(1-z)^{q_1+q_4-q_2-q_3})[q_3q_4q_1q_2],\\
	(14)(23)\,&:\,[q_1q_2q_3q_4]\mapsto n((-z)^{q_3+q_4-q_1-q_2})[q_4q_3q_2q_1].
}
Here $n(x)=x/\sqrt{x\bar x}$, where $\bar x$ is $x$ with $z$ and $\bar z$ exchanged. The possible $(-1)$'s from permutations of fermions are already taken into account. Note that if a structure is fixed by a permutation, the phase factor is automatically $1$. This is due to the hidden triviality of these phases mentioned in the appendix~\ref{app:permutations}. This means that any structure can be symmetrized to give a non-zero result, 
\beq
\label{eq:symmetrization3dfourpoints}
	\crr{q_1q_2q_3q_4}_z=\frac{1}{n_{q_1q_2q_3q_4}}\sum_{\pi\in\Pi^\text{kin}}\pi[q_1q_2q_3q_4]\neq 0,
\eeq
where $n_{q_1q_2q_3q_4}$ is the number of elements in $\Pi^\text{kin}$ stabilizing $[q_1q_2q_3q_4]$.
With this notation a $\Pi^\text{kin}$-symmetric four-point function can be rewritten as
\beq
	g_0(s_i,z,\bar z)=\sum_{q_i/\Pi^{\text{kin}}}\crr{q_1q_2q_3q_4}_z g_{[q_1q_2q_3q_4]}(z,\bar z),
\eeq
where the sum is over some set of representatives of orbits of $\Pi^\text{kin}$ action on the set of all tensor structures (possibly of definite parity).

For four-point functions it is convenient to also consider the action of the permutation $(13)$, which is often used to write down a bootstrap equation for a four-point function containing identical operators. From the results of appendix~\ref{app:permutations}, it acts as
\beq
	(13)\,:\,[q_1q_2q_3q_4]\mapsto (-1)^{q_1+q_2-q_3-q_4}[q_3q_2q_1q_4],
\eeq
and this already accounts for the $(-1)$ sign coming from a possible permutation of fermions.
For the symmetrized structures the action is, including the change $z\to 1-z$ 
\beq
	\crr{q_1q_2q_3q_4}_z\mapsto(-1)^{q_1+q_2-q_3-q_4}\crr{q_3q_2q_1q_4}_z.
\eeq
The crossing equation for the full four-point function, in the case when the operators $1$ and $3$ are identical, is
\beq
\label{eq:crossing3d}
\sum_{q_i/\Pi^{\text{kin}}}\crr{q_1q_2q_3q_4}_z g_{[q_1q_2q_3q_4]}(z,\bar z)=\sum_{q_i/\Pi^{\text{kin}}}\crr{q_3q_2q_1q_4}_z (-1)^{q_1+q_2-q_3-q_4}g_{[q_1q_2q_3q_4]}(1-z,1-\bar z).
\eeq
Note that the crossing permutation $(13)$ maps orbits of $\Pi^\text{kin}$ into orbits, so this basis essentially diagonalizes the crossing equation.

\paragraph{Counting.} It is easy to count the number of four-point structures. Clearly, the total number of structures is
\beq
\label{eq:counting3dfourpoint}
	N_{3d}(\ell_1,\ell_2,\ell_3,\ell_4)=\prod_{i=1}^{4}(2\ell_i+1),
\eeq
and as discussed in section~\ref{sec:npointconformalframe}, this result is valid for all higher-point functions,
\beq
	N_{3d}(\ell_1\ldots \ell_n)=\prod_{i=1}^n(2\ell_i+1),\quad n\geq 4.
\eeq

One can see from~\eqref{eq:parity3dfourpoint} that if there is at least one half-integer spin, then the number of parity even structures is equal to the number of parity odd structures (for such a spin $\ell_i-q_i$ is even exactly as often as it is odd). Performing an explicit computation in the case when all spins are integral, we arrive at the direct analog of~\eqref{eq:3ptnparity}
\beq
\label{eq:counting3dfourpointparity}
N^\pm_{3d}(\ell_1,\ell_2,\ell_3,\ell_4)=\frac{N_{3d}(\ell_1,\ell_2,\ell_3,\ell_4)\pm\kappa}{2},
\eeq
where $\kappa=1$ when all spins are integral and $\kappa=0$ otherwise.

If there are non-trivial kinematic permutations, these are $\Pi^\text{kin}=\mathbb{Z}_2$ or $\Pi^\text{kin}=\mathbb{Z}_2^2$. In each case we can either use proposition~\ref{prop:fourptperms} and~\eqref{eq:characterz2z2} or count the number of orbits of $\Pi^\text{kin}$ action on $[q_1q_2q_3q_4]$ structures, which can be done using Burnside's lemma. The result in $\mathbb{Z}_2$ case is
\al{
	N^+_{3d}(\ell_1\leftrightarrow\ell_2,\ell_3\leftrightarrow\ell_4)&=\frac{1}{2}\left[N^+_{3d}(\ell_1,\ell_1,\ell_3,\ell_3)+(2\ell_1+1)(2\ell_3+1)\right],\\
	N^-_{3d}(\ell_1\leftrightarrow\ell_2,\ell_3\leftrightarrow\ell_4)&=\frac{1}{2}N^-_{3d}(\ell_1,\ell_1,\ell_3,\ell_3).
}
The result in $\mathbb{Z}_2^2$ case is 
\al{
\label{eq:counting3dfourpointevenz2sq}
	N^+_{3d}(\ell)&=\frac{1}{4}\left[N^+_{3d}(\ell,\ell,\ell,\ell)+3(2\ell+1)^2\right],\\
	N^-_{3d}(\ell)&=\frac{1}{4}N^-_{3d}(\ell,\ell,\ell,\ell).
}

\subsection{Example: 4 Majorana fermions}
\label{sec:3d4Majorana}

As an example, let us consider in detail the case of four identical Majorana fermions. This is a relatively simple yet non-trivial case for which we can compare to~\cite{Iliesiu:2015qra}.

Let us start by analyzing the generic three-point functions for operators which appear in the OPE expansion. First, consider the three point function of two distinct Majorana fermions and a spin-$\ell_3$ operator. Using \eqref{eq:3dthreepointparity} and \eqref{eq:3dthreepointneutrality}, we find the following structures,
\al{
[\tfrac{1}{2},\tfrac{1}{2},-1]^\pm,\quad
[\tfrac{1}{2},-\tfrac{1}{2},0]^\pm.
}
For $\ell_3=0$ we can only have $q_3=0$, and thus only $1$ parity-even and $1$ parity-odd structures remain.
If the fermions are identical, then we need only the structures symmetric under the exchange $(12)'$ given by~\eqref{eq:3perm12f}. This leaves for even $\ell_3$ 
\al{
	[\tfrac{1}{2},\tfrac{1}{2},-1]^+,\quad 
	[\tfrac{1}{2},-\tfrac{1}{2},0]^\pm,
}
and for odd $\ell_3$
\al{
	&[\tfrac{1}{2},\tfrac{1}{2},-1]^-.	
}
This is in complete agreement with~\cite{Iliesiu:2015qra}.

Let us now turn to four-point functions. First, using~\eqref{eq:counting3dfourpoint}, we immediately find that there are $2^4=16$ tensor structures. According to~\eqref{eq:counting3dfourpointparity}, $8$ of them are parity-even and $8$ are parity-odd. Using~\eqref{eq:parity3dfourpoint} we can write down the parity-even structures, denoting $q=+\tfrac{1}{2}$ with $\up$ and $q=-\tfrac{1}{2}$ with $\down$,
\al{
\begin{split}
	&[\up\up\up\up],[\down\down\down\down], \\
	&[\up\up\down\down],[\down\down\up\up],\\
	&[\up\down\up\down],[\down\up\down\up],\\
	&[\up\down\down\up],[\down\up\up\down].
\end{split}
}
Assuming that the fermions are identical, we simply perform the $\mathbb{Z}_2^2$ symmetrization~\eqref{eq:symmetrization3dfourpoints} of these structures, obtaining $5=(8+3\cdot 2^2)/4$ (c.f.~\eqref{eq:counting3dfourpointevenz2sq}) independent parity-even structures,
\al{
	&\crr{\up\up\up\up},\crr{\up\up\down\down},\crr{\up\down\up\down},\crr{\down\up\up\down},\crr{\down\down\down\down}.
}
We can also easily form crossing-symmetric and anti-symmetric structures using~\eqref{eq:crossing3d},
\al{
\label{eq:fourmajoranacrossingeven}
	\text{symmetric: }&\crr{\up\up\up\up},\crr{\up\down\up\down},\crr{\down\down\down\down},\crr{\up\up\down\down}+\crr{\down\up\up\down},\\
\label{eq:fourmajoranacrossingodd}
	\text{anti-symmetric: }&\crr{\up\up\down\down}-\crr{\down\up\up\down}.
}
We thus have 4 crossing-even structures and 1 crossing-odd structure, which lead to 4 crossing-even equations and 1 crossing-odd equation.\footnote{Note however that ``crossing parity'' is not a real invariant and can be modified by a structure redefinition.} This again coincides with the results of~\cite{Iliesiu:2015qra}. 

We can very explicitly write down the standard basis of crossing equations,
\al{
\label{eq:fourmajoranacrossingderiveven}
	\partial^n\bar\partial^m g_{[\up\up\up\up]}=
	\partial^n\bar\partial^m g_{[\down\down\down\down]}=
	\partial^n\bar\partial^m g_{[\up\down\up\down]}=
	\partial^n\bar\partial^m (g_{[\up\up\down\down]}+ g_{[\down\up\up\down]})=0,&\quad n+m\text{ odd},\\
\label{eq:fourmajoranacrossingderivodd}
	\partial^n\bar\partial^m (g_{[\up\up\down\down]}- g_{[\down\up\up\down]})=0,&\quad n+m\text{ even},
}
where all functions are evaluated at $z=\bar z=1/2$. However, there is an important subtlety. When we expand the four-point function in conformal blocks, we will find that the result is smooth (as a function of $x_i$). As we discuss in appendix~\ref{sec:smoothness}, not any choice of $g_{[q_1q_2q_3q_4]}(z,\bar z)$ leads to a smooth correlator, and a finite number of boundary conditions need to be imposed on derivatives of $g_{[q_1q_2q_3q_4]}(z,\bar z)$ at $z=\bar z$. This effectively gives relations between equations~\eqref{eq:fourmajoranacrossingderiveven} and~\eqref{eq:fourmajoranacrossingderivodd}. These are easy to classify, and we work out the present example in appendix~\ref{sec:smoothness}.

Note that~\cite{Iliesiu:2015qra} used 4-point tensor structures constructed using embedding-space building blocks. They did not have to perform the aforementioned analysis of the boundary conditions. However, there was a different problem which required a similar analysis --- since their coefficient functions, unlike those in the present work, do not represent physical values of the correlator but rather have to be multiplied by their tensor structures first, it is not guaranteed that they do not have singularities. In fact, it was found in~\cite{Iliesiu:2015qra} that their coefficient functions for conformal blocks diverge as $(z-\bar z)^{-5}$ near $z=\bar z$. The solution was to multiply these functions by $(z-\bar z)^5$ at the cost of introducing relations between the Taylor series coefficients, which are similar to ours. What is different is that in our case we have a simple classification of these relations, whereas in~\cite{Iliesiu:2015qra} they were handled in a brute-force way by numerically finding linearly independent vectors of crossing equations.

\section{Scattering amplitudes}
\label{sec:scattering}

In this section we establish the equivalence of the counting of conformal correlators in CFT$_d$ with counting of scattering amplitudes in flat space QFT$_{d+1}$. The basic idea is quite simple --- the conformal frame approach can be applied to scattering amplitudes in QFT$_{d+1}$, and it yields equivalent group-theoretic formulas.

Let us formulate the counting problem for amplitudes in the simplest case of traceless-symmetric spin $\ell$ particles (we will generalize to other representations later in this section). We can describe the scattering amplitude $\Amp(p_i,\zeta_i)$ as a Lorentz-invariant function of the momenta $p_i$, $p_i^2=-m_i^2$, $\sum_i p_i=0$, and traceless symmetric polarizations $\zeta_i^{\mu_1\ldots\mu_{\ell_i}}$. For all particles the polarizations satisfy the transversality condition $(p_i)_{\mu_1}\zeta_i^{\mu_1\ldots\mu_{\ell_i}}=0$. For massless particles we in addition get the gauge equivalence
\beq
	\zeta^{\mu_1\ldots\mu_{\ell_i}}\sim \zeta^{\mu_1\ldots\mu_{\ell_i}}+p^{(\mu_1}\lambda^{\mu_2\ldots\mu_\ell)},
\eeq
where $\lambda$ is the parameter of the gauge transformation which is itself transverse.
The scattering amplitude $\Amp(p_i,\zeta_i)$ should be invariant under this transformation. That is, $\Amp$ should be a function of the gauge equivalence classes of $\zeta_i$. 

A general solution to the above requirements has the form
\beq
	\Amp(p_i,\zeta_i)=\sum_{I=1}^{N}\mathbb{T}_I(p_i,\zeta_i)f^I(s,t,\ldots),
\eeq
where $\mathbb{T}_I$ are the tensor structures encoding the non-trivial dependence on the polarizations and momenta, and $s,t,\ldots$ are the kinematic invariants of $n$ particles, i.e.\ the Mandelstam variables. Our goal in this section is to find the number $N$ of tensor structures and prove that it is equal to the number of tensor structures in a certain conformal correlator.

\subsection{Little group formulation}
Note that for a fixed $p$, the solutions $\zeta$ to the transversality constraint $p_{\mu_1}\zeta^{\mu_1\ldots\mu_{\ell}}=0$, as well as the gauge equivalence classes of such solutions are transformed into each other by the little group $L(p)$ which is the subgroup of the Lorentz group leaving $p$ invariant. The little group in QFT$_{d+1}$ is $SO(d)$ in the massive case and $SO(d-1)$ in the massless case (formally it is $ISO(d-1)$, but for particles with a finite number of internal degrees of freedom the translations of $ISO$ act trivially).  In the case considered above $\zeta_i$ live in traceless symmetric representations of the respective little groups. 

In order to have a general treatment, we will adopt this little group point of view on the particle polarizations. Instead of specifying a polarization $\zeta$, we specify an element $\varepsilon$ of some representation of $L(k)$, where $k$ is a standard\footnote{For concreteness, for massive particles of mass $m$ we can choose $k=(m,0,0,\ldots)$ and for massless particles $k=(1,1,0,0,\ldots)$ with signature $(-,+,+,\ldots)$.} momentum with $k^2=p^2$. Accordingly, for each momentum $p$ we specify a standard Lorentz transformation\footnote{In general we need to allow $R(p)$ to belong to the disconnected components of the Lorentz group, since in general we may want to have momenta in the past lightcone (or treat in and out particles separately). Alternatively, we may consider the complexification of the whole setup, as anyway is required for the treatment of $3$-point on-shell amplitudes. Either way, for simplicity of the discussion we ignore these subtleties.} $R(p)$ such that $R(p)k=p$. Now instead of $\Amp(p_i,\zeta_i)$, we have a function of the little group polarizations $\varepsilon_i$ which we denote $S(p_i,\varepsilon_i)$. 

To see the correspondence between the two descriptions, for example in the case of massless traceless symmetric particle, we can put $\varepsilon$ into correspondence with a polarization $\zeta_k(\varepsilon)$ with transversality and gauge invariance defined by the momentum $k$. This then specifies $\zeta_p(\varepsilon)=R(p)\zeta_k(\varepsilon)$, which now satisfies transversality and gauge invariance defined by $p$. We can now set 
\beq
S(p_i,\varepsilon_i)=\Amp(p_i,\zeta_{p_i}(\varepsilon_i)).
\eeq 
This establishes the isomorphism between the descriptions $S(p_i,\varepsilon_i)$ and $\Amp(p_i,\zeta_i)$. It also makes it easy to see how the Lorentz invariance is stated for $S(p_i,\varepsilon_i)$ --- since for each Lorentz transformation $\Lambda$ we have
\beq
	\Amp(\Lambda p_i,\Lambda \zeta_i)=\Amp(p_i,\zeta_i),\label{eq:amplorentz}
\eeq
then in terms of $S(p_i,\varepsilon_i)$ we should have
\beq
	S(\Lambda p_i,R(\Lambda p_i)^{-1}\Lambda R(p_i)\varepsilon_i)=S(p_i,\varepsilon_i).
\eeq
This formula makes sense because $R(\Lambda p_i)^{-1}\Lambda R(p_i)k_i=k_i$ and thus $R(\Lambda p_i)^{-1}\Lambda R(p_i)\in L(k_i)$, which can act on $\varepsilon_i$. This condition appears more complicated than~\eqref{eq:amplorentz}, but the advantage is that this is the only condition we require of the amplitude (in contrast to requiring the gauge invariance and imposing the transversality constraints for $\Amp(p_i,\zeta_i)$). This makes it extremely easy to classify tensor structures for the amplitudes, as we now show.

\subsection{Conformal frame for amplitudes}

We now simply repeat the analysis of section~\ref{sec:conformalframe} for the amplitudes. The Lorentz group acts on the configuration space of the momenta $p_i$, and splits this space into orbits. We chose a ``scattering frame'' --- a submanifold of the momenta configuration space which intersects each orbit at precisely one point. It is easy to show that the dimension of scattering frame is the same as the dimension of the conformal frame at the same $n$ (number of operators or particles) and $d$.

A scattering amplitude is now completely specified by its values on the scattering frame. These values, as in section~\ref{sec:conformalframe}, have to be invariant under the subgroup of Lorentz group which fixes the scattering frame. 

It is easy to see what this subgroup is. First, $n$ generic momenta, due to the conservation condition $\sum_i p_i=0$, span an $(m-1)$-dimensional linear space $\PP$, where $m=\min(d+2,n)$. The subgroup which fixes $\PP$ depends only on the rank of the restriction of the Lorentz metric onto $\PP$, which coincides with the rank of Gram matrix $G$ of any $n-1$ momenta in $\PP$. The determinant $\det G$ is an algebraic function of the particle masses $m_i$ and the kinematic invariants $s,t,u,\ldots$.

For $n\geq 4$ we have non-trivial kinematic invariants, and thus for a generic set of these invariants $\det G\neq 0$ and the metric on $\PP$ is full rank. This implies that $\PP$ is stabilized by a subgroup $SO(d+1-(m-1))=SO(d+2-m)$.

For $n=3$\footnote{For $n=3$ we need to consider complexified kinematics in order to have an on-shell amplitude.} we have no non-trivial kinematic invariants, and $\det G$ is determined solely by the masses. For a generic set of masses, $\det G\neq 0$, and we again get $SO(d+2-m)$. This case corresponds to the generic three-point functions as discussed in section~\ref{sec:conserved}. For simplicity, we only consider this generic case. We comment on the non-generic case in future work~\cite{KravchukFuture}.

Now, we need to understand how the stabilizing subgroup $St=SO(d+2-m)$ acts on the little group polarizations. Assume that $\Lambda$ fixes all the $p_i$. In this case, we have
\beq
	\varepsilon_i\to R(p_i)^{-1}\Lambda R(p_i)\varepsilon.
\eeq
We can say, alternatively, that $St$ is naturally a subgroup of each $L(p_i)$, which in turn are put in an isomorphism with $L(k_i)$ by
\beq
	L(k_i)=R(p_i)^{-1}L(p) R(p_i).
\eeq
This defines a restriction of representations of $L(k_i)$ to representations of $St=SO(d+2-m)$. Assume that the particles transform in representations $\rho_i$ of $L(k_i)$. We then immediately find that the space of tensor structures for scattering amplitudes is
\beq
	\p{\bigotimes_{i=1}^n \mathrm{Res}^{L(k_i)}_{SO(d+2-m)}\rho_i}^{SO(d+2-m)}.
\eeq
Its dimension is equal to the number of tensor structures in a conformal correlator if the $SO(d)$ representations of the non-conserved local operators in CFT$_d$ are identified\footnote{This rule for massive case was also mentioned in~\cite{Costa:2011mg,Costa:2014rya}.} with the representations of the massive little group $SO(d)$ in QFT$_{d+1}$, and the effective $SO(d-1)$ representations of local operators (as described in section~\ref{sec:conserved}) are identified with the representations of the massless $SO(d-1)$ little group. It is in principle straightforward to extend this result to include parity and permutations symmetries. For example, it is not hard to check that kinematic permutation groups match in CFT$_d$ and QFT$_{d+1}$.

\paragraph{Note added:} When this paper was being prepared, the work~\cite{Schomerus:2016epl} appeared, which gives the formula~\eqref{eq:3ptbasic} and its generalization to the four point function case. However, in~\cite{Schomerus:2016epl} the questions of actual construction of tensor structures, parity and permutation symmetries, and conservation conditions were not considered.

\section*{Acknowledgements}

We thank Chris Beem, Clay C\'ordova, Tolya Dymarsky, Mikhail Evtikhiev, Enrico Herrmann, Denis Karateev, Hirosi Ooguri and Chia-Hsien Shen for discussions. DSD is supported by a William D. Loughlin Membership at the Institute for Advanced Study, and Simons Foundation grant 488657 (Simons Collaboration on the Non-perturbative Bootstrap). This material is based upon work supported by the U.S. Department of Energy, Office of Science, Office of High Energy Physics, under Award Numbers DE-SC0009988~(DSD) and DE-SC0011632~(PK).

\pagebreak
\appendix

\section{Smoothness conditions on correlators}
\label{sec:smoothness}
The analysis of section~\ref{sec:conformalcorrelators} did not take into account smoothness of $g$. In order for $g$ to be continuous, it is sufficient for $g_0$ to be continuous and to satisfy the stabilizer invariance condition~\eqref{eq:confinvCF}. Note that with the choice of conformal frame discussed in section~\ref{sec:npointconformalframe} the stabilizer subgroup is the same $SO(d-m+2)$ for generic $y$, but it enhances an the boundaries of conformal frame, essentially giving a boundary condition for the otherwise $SO(d+m-2)$-invariant $g_0$. We will now see that this boundary condition needs to be refined further if we want $g$ to be smooth.

For simplicity, let us consider only the most important case of $4$-point functions. It is easy to convince oneself that $g$ as given by~\eqref{eq:cfreconstruction} will be smooth for $y$ in the interior of conformal frame as soon as $g_0$ is smooth there. What is non-trivial is the smoothness on the boundary of conformal frame. Let us start with a smooth $g$ and see what kind of $g_0$ it leads to. 

We split the reduction to conformal frame into two steps. First, we fix the coordinates $x_1,x_3,x_4$ as in section~\ref{sec:conformalframe}. This leads to a function $g_1(x_2)$ which is to be invariant under $SO(d-1)$. Note that its smoothness is equivalent to smoothness of $g$. We can expand $g_1$ in Taylor series along the directions orthogonal to $e$,
\beq
	g_1(\pol_i,x_2)=\sum_{n=0}^{N}g_1^{\mu_1\ldots\mu_n}(\pol_i,e\cdot x_2)z_{\mu_1}\ldots z_{\mu_n}+o(z^N),
\eeq
where $z$ is the $(d-1)$-dimensional projection of $x_2-e$ onto the subspace orthogonal to $e$. From the invariance equation~\eqref{eq:polarizationinvariance} we read off the condition that for every $e\cdot x_2$ the value $g_1^{\mu_1\ldots \mu_n}(\,\cdot\,,e\cdot x_2)$ is a singlet in
\beq
	\hat{\mathbf{n}}\otimes\mathrm{Res}^{O(d)}_{O(d-1)}\bigotimes_{i=1}^{4}\rho_i,\label{eq:taylorselection}
\eeq
where $\hat{\mathbf{n}}$ is the reducible symmetric tensor representation of $O(d-1)$, and we included the possibility of permutation\footnote{We can analyze permutations by using conformal transformations in the plane spanned by $x_2$ and $e$.} and parity symmetries. The symmetric tensor decomposes into symmetric traceless tensors as
\beq
	\hat{\mathbf{n}} = \mathbf{n}+(\mathbf{n-2})+\ldots + (\mathbf{n}\!\!\mod 2).\label{eq:tracefullexpansion}
\eeq

Now when we finally restrict to the conformal frame by taking $z$ inside the half-plane $\alpha$, which we will assume to be along 1st and 2nd coordinate axes, with $e$ being along the 1st axis, we find
\beq
\label{eq:smoothnessseries}
	g_0(\pol_i,x_2^1,x_2^2)=\sum_{n=0}^{N}g_1^{2\ldots2}(\pol_i, x_2^1)(x_2^2)^n+o\left((x_2^2)^N\right).
\eeq
Note that theorem~\ref{thm:main} tells us to look for $O(d-1)$ symmetric traceless tensors\footnote{This is equivalent to taking singlets in further restriction to $O(d-2)$} in
\beq
\label{eq:smoothnesstensorproduct}
\mathrm{Res}^{O(d)}_{O(d-1)}\bigotimes_{i=1}^{4}\rho_i.
\eeq
Equation~\eqref{eq:taylorselection} therefore tells us at which orders in Taylor series~\eqref{eq:smoothnessseries} which traceless symmetric tensors of~\eqref{eq:smoothnesstensorproduct} can contribute. For example, the spin-$3$ symmetric traceless tensor representation $\mathbf{3}$, if appears in~\eqref{eq:smoothnesstensorproduct}, defines a tensor structure whose coefficient function can contribute to~\eqref{eq:smoothnessseries} at orders $(x_2^2)^3,(x_2^2)^5,(x_2^2)^7,\ldots$ but not $(x_2^2)^1$ or $(x_2^2)^{2n}$.

In other words,~\eqref{eq:taylorselection} restricts the expansion of the coefficient functions of our structures by specifying their parity under $x_2^2\to -x_2^2$ and the rate at which they go to zero on the boundary of conformal frame. Note that the $x_2^2$ parity of the coefficient function can also be extracted from how the corresponding structure behaves under a $\pi$ rotation in the plane, say, $2$-$3$, which is more convenient in practice than~\eqref{eq:taylorselection}.

As the most basic example, consider the scalar four-point function. In this case, the Taylor coefficients $g_1^{\mu_1\ldots \mu_n}$ are singlets in 
\beq
	\hat{\mathbf{n}}\otimes \bullet=\hat{\mathbf{n}},
\eeq
and thus only exist for even $n$, according to~\eqref{eq:tracefullexpansion}. This tells us that scalar correlation functions restrict to $g_0$ with even expansion in $x_2^2$ and this is why we can parametrize them by $u$ and $v$ (which are also even). 

\subsection{Example: 4 Majorana fermions}

Consider now the example of section~\ref{sec:3d4Majorana}. 
There are two aspects of the smoothness analysis which are important for actual numerical analysis. For convenience, we use the $t$ and $x$ coordinates of section~\ref{sec:3d4pt} below.

The first is that some of the coefficient functions are restricted to be even or odd in $t\sim z-\bar z$. This is easy to handle by hand, since as noted above, this is determined by the behavior of the structure under $\pi$ rotation in the plane $0$-$2$. Via analytic continuation this rotation is equivalent to exchange of $\up$ and $\down$. Therefore, we can consider structures
\al{
	\left\langle\up\up\up\up\right\rangle^\pm&=\left\langle\up\up\up\up\right\rangle \pm\left\langle\down\down\down\down\right\rangle,\nn\\
	\left\langle\up\up\down\down\right\rangle^+&=\left\langle\up\up\down\down\right\rangle+\left\langle\down\down\up\up\right\rangle,\nn\\
	\left\langle\up\down\up\down\right\rangle^+&=\left\langle\up\down\up\down\right\rangle+\left\langle\down\up\down\up\right\rangle,\nn\\
	\left\langle\down\up\up\down\right\rangle^+&=\left\langle\up\down\down\up\right\rangle+\left\langle\down\up\up\down\right\rangle,
}
each of which have definite parity under $t\to -t$. Note that we didn't form the difference in the last three structures since the terms on the right side in each line lie in the same orbit of $\mathbb{Z}_2^2$.

The second is that some of the coefficient functions should vanish faster than is required by their $t$-parity. We compute\footnote{In general one may need to be a little more careful with the permutation phases than we have been in this simple example.}, using~\eqref{eq:z2sqprescription}
\beq
	\mathrm{Res}^{O(3)}_{O(2)}\p{\mathbf{\tfrac{1}{2}}^{\otimes 4}}^{\mathbb{Z}_2^2}=\mathbf{2}\oplus \mathbf{1}\oplus 3\,\bullet^+.
\eeq
According to~\eqref{eq:taylorselection}, this means that from $5$ coefficient functions of parity-even structures, 4 are even in $t$, of which $3$ start with $t^0$ and $1$ starts with $t^2$, and one is odd in $t$ and starts with $t^1$. We see that there is one $t$-even coefficient function which should vanish as $t^2$, which is faster than required by its $t$-parity. This means that there is a linear relation between $t^0$ coefficients of the coefficient functions $g_{\langle\up\up\up\up\rangle^+},g_{\langle\up\up\down\down\rangle^+},g_{\langle\up\down\up\down\rangle^+},g_{\langle\down\up\up\down\rangle^+}$, i.e.\
\beq
	\alpha_1 g_{\langle\up\up\up\up\rangle^+}(0,x)+\alpha_2 g_{\langle\up\up\down\down\rangle^+}(0,x)+\alpha_3 g_{\langle\up\down\up\down\rangle^+}(0,x)+\alpha_4 g_{\langle\down\up\up\down\rangle^+}(0,x)=0,
\eeq
where the first argument is $t=0$. One can check that $\alpha_1\neq 0$, and we can then use this equation to find $g_{\langle\up\up\up\up\rangle^+}(0,x)$. More generally, to find such relations, one can consider the action of the $e$-preserving $SO(d-1)$ rotation generators on the tensor structures to decompose these structures according to $SO(d-1)$ and then use~\eqref{eq:taylorselection}.

Summarizing the discussion in section~\ref{sec:3d4Majorana} and in this appendix, one can use the following independent system of crossing equations,
\al{
	\partial_t^{2n}\partial_x^{2m+1} g_{\langle\up\up\up\up\rangle^+}=0,&\quad n\geq 1,m\geq 0,\nn\\
	\partial_t^{2n}\partial_x^{2m+1} g_{\langle\up\down\up\down\rangle^+}=0,&\quad n\geq 0,m\geq 0,\nn\\
	\partial_t^{2n}\partial_x^{2m+1} \left(g_{\langle\up\up\down\down\rangle^+}+g_{\langle\down\up\up\down\rangle^+}\right)=0,&\quad n\geq 0,m\geq 0,\nn\\
	\partial_t^{2n}\partial_x^{2m} \left(g_{\langle\up\up\down\down\rangle^+}-g_{\langle\down\up\up\down\rangle^+}\right)=0,&\quad n\geq 0,m\geq 0,\nn\\
	\partial_t^{2n+1}\partial_x^{2m} g_{\langle\up\up\up\up\rangle^-}=0,&\quad n\geq 0,m\geq 0,
}
where everything is evaluated at $t=0,\,x=1/2$.

\section{More on permutations}
\label{app:permutations}

\subsection{Kinematic permutations}
\label{app:permutationskinetic}

In this section we prove that $\{S^\text{kin}_n\}_{n=1}^\infty=\{0,S_2,S_3,\mathbb{Z}_2^2,0,0,\ldots\}$, where $0$ stands for the trivial group.  The first three cases are, as noted in the main text, trivial, since the conformal moduli space $\overline \MM_n$ of $n=1,2,3$ points consists of one point, and thus $S_n^\text{kin}=S_n$.

Now suppose $n\geq 4$. Consider the set $U$ of all conformal cross-ratios of the form 
\beq
	u_{ij,kl}=\frac{x_{ij}^2 x_{kl}^2}{x_{ik}^2 x_{jl}^2},\quad x_{ij}^2=(x_i-x_j)^2,
\eeq
with $i,j,k,l$ all different.
Permutations of points $x_i$ act on this set by permutations, and permutations from $S_n^\text{kin}$ should leave these cross-ratios invariant. Since for a generic configuration there are no two exactly equal cross-ratios (even though there are relations between them), this means that the permutations induced on $U$ should be trivial.

Suppose a permutation maps $i\to j,\, i\neq j$. Then by looking at $u_{ij,kl}$ (with $i,j,k,l$ all different) we see that necessarily $j\to i$, otherwise this cross-ratio will change. But then also $k\leftrightarrow l$. Since we were free to choose $k,l$, this leads to a contradiction unless $n=4$ and only one choice of $k,l$ is possible. This establishes that $S_n^\text{kin}=0$ for $n>4$. For $n=4$ it means that the allowed permutations are products of 2-cycles and an explicit check shows that all possible products are allowed, giving
\beq
	S_4^\text{kin}=\{e,(12)(34),(13)(24),(14)(23)\}=\mathbb{Z}_2^2.
\eeq

\subsection{Conformal transformations for permutations}
\label{app:permutationsphases}
We now analyze explicitly the conformal transformations $r_\pi$ induced by permutations. We only do so for three and four-point functions, since these are the only cases when there are interesting kinematic permutations.

For both three- and four-point functions we choose the $r_\pi$ to preserve the plane $\alpha$ in which all the operators lie (for three points we choose some such plane). Such conformal transformations restrict on $\alpha$ to the fractional linear transformations, and we can describe them by a mapping
\beq
	x\mapsto x'=\frac{ax+b}{cx+d},
\eeq
where we identified $\alpha$ with $\mathbb{C}$.
Note that we can choose these transformation to give trivial rotations in the planes orthogonal to $\alpha$. We therefore only need to compute $Spin(2)$ elements induced by $r_\pi$ inside the plane, and the problem is entirely two-dimensional.

The group of fractional linear transformations is double covered by $SL(2,\mathbb{C})$. Thus $r_\pi\in SL(2,\mathbb{C})$. The correspondence is
\beq
	r_\pi=\begin{pmatrix}
	a & b \\ c & d
	\end{pmatrix}\in SL(2,\mathbb{C})\Rightarrow r_\pi x = \frac{ax+b}{cx+d},\quad ad-bc=1.
\eeq
This is 2 to 1 because $r_\pi$ and $-r_\pi$ give the same transformation. Recall that the basic condition for $r_\pi$ is that 
\beq
	r_\pi x'_i=x_{\pi(i)},
\eeq
for $x'_i$ in the conformal frame. In the case of kinematic permutations we have $x'_i=x_i$. Thus we have the following equation for $r_\pi$,
\beq
	\frac{ax_i+b}{cx_i+d}=x_{\pi(i)}.
\eeq
This has two solutions differing by a sign. Since the correlator is bosonic in total, we are free to choose either of them.

The $SO(2)$ element $R_{r_\pi}(x_i)$ is given by (upon identification of $SO(2)$ with the unit circle in complex plane)
\beq
	R_{r_\pi}(x_i)=\left.n\left(\frac{dx'}{dx}\right)\right\vert_{x=x_{\pi^{-1}(i)}},
\eeq
where $n(x)=x/|x|$. The implementation of the lifting from $SO(2)$ to $Spin(2)$ discussed in section~\ref{sec:conformalinvariance} is straightforward in two dimensions. Note that for $ad-bc=1$,
\beq
	\frac{d}{dz}\left(\frac{ax+b}{cx+d}\right)=\frac{1}{(c z +d)^2}.
\eeq
This is invariant under $r_\pi\to -r_\pi$ and the phase gives an element of $SO(2)$ as above. Lifting to an element of $Spin(2)$ is essentially equivalent to choosing a square root of this expression, with the most natural choice being
\beq
	\sqrt{\frac{d}{dx}\left(\frac{ax+b}{cx+d}\right)}=\frac{1}{c x +d}.
\eeq
This is not invariant over $r_\pi\to-r_\pi$, which means that this is only a map from the double cover $SL(2,\mathbb{C})$ of the conformal group to $Spin(2)$, but not from the conformal group $PSL(2,\mathbb{C})=SO(3,1)$ itself. This is in accord with the discussion in section~\ref{sec:conformalinvariance}. Therefore, we find that 
\beq
	\RR_{r_\pi}(x_i)=n(c x+d)^{-1}\vert_{x=x_{\pi^{-1}(i)}}.
\eeq

In the following table we summarize the locations of the operators in the conformal frame we choose, by specifying the complex coordinates
\begin{center}
\begin{tabular}{r||c|c|c|c}
 & $x_1$ & $x_2$ & $x_3$ & $x_4$ \\
\hline
\hline
3-point & 0 & 1 & $\infty$ & - \\
\hline
4-point & 0 & $z$ & 1 & $\infty$
\end{tabular}
\end{center}

As discussed before, the operator at infinity is inserted by putting it at $L$ and then taking the limit $L\to \infty$ along the real axis. This is done in order to avoid using inversion when defining the operator at infinity. A safe way of determining the phases is working with finite $L$ and then taking the limit.

In the following we compute the transformations  $r_\pi$ and
\beq
	\RR_{r_\pi}(x_i)^{-1}=n(h_i(\pi)).
\eeq
Note that the $SO(2)$ rotation angle is given by the phase of $n(h_i(\pi))^2$. We write the permutations in cycle notation. For example, $\pi=(134)(25)$ is the permutation $\pi(1)=3,\pi(3)=4,\pi(4)=1,\pi(2)=5,\pi(5)=2$.

\subsubsection{3-point functions}
\label{app:permutations3pt}

For three-point functions we have the following parameters $a,b,c,d$ for the transformations and the induced $h_i$:

\begin{center}
\begin{tabular}{r || c | c | c || c | c | c | c}
		& $h_1$ 	& $h_2$ 	& $h_3$ & $a$ & $b$ & $c$ & $d$ \\\hline\hline
id			& $1$ 	& $1$ 	& $1$	&1&0&0&1	\\\hline
(12)			& $i$	& $i$	& $-i$ &$-1$&1&0&1	\\\hline
(13)			& $i$	& $i$	& $i$ &0&1&1&0	\\\hline
(23)			& $-i$	& $i$ 	& $i$	&1&0&1&$-1$\\\hline
(123)		& $-1$	& $1$	& $1$	&0&1&$-1$&1	\\\hline
(132)		& $1$	& $1$	& $-1$ &1&-1&1&0
\end{tabular}
\end{center}

\subsubsection{4-point functions}
\label{app:permutations4pt}

For four-point functions we have 
\begin{center}
\begin{tabular}{r || c | c | c | c }
			& $h_1$ 		& $h_2$ 		& $h_3$ 		& $h_4$ 		\\
\hline\hline
id			& $1$ 		& $1$ 		& $1$		&	 $1$			\\\hline
$(12)(34)$	& $i\sqrt{1-z}$	& $i\sqrt{1-\bar z}$		& $-i\sqrt{1-\bar z}$ 	&	$-i\sqrt{1-z}$	\\\hline
$(13)(24)$	& $-\sqrt{\bar z(1-z)}$		& $-\sqrt{\bar z(1-\bar z)}$		& $\sqrt{z(1-\bar z)}$ 		&	$\sqrt{z(1-z)}$			\\\hline
$(14)(23)$	& $i\sqrt{\bar z}$		& $i\sqrt{\bar z}$ 		& $i\sqrt{z}$		&	$i\sqrt{z}$		\\\hline
\end{tabular}
\end{center}

Note that these transformations have to be accompanied by a $-$ sign for an odd permutation of fermions. If we assume that we use the permutations to exchange identical operators then we can instead use the following table, but \textit{without} the extra minus sign for the odd fermion permutation,
\begin{center}
\begin{tabular}{r || c | c | c | c }
			& $\tilde h_1$ 		& $\tilde h_2$ 		& $\tilde h_3$ 		& $\tilde h_4$ 		\\
\hline\hline
id			& $1$ 		& $1$ 		& $1$		&	 $1$			\\\hline
$(12)(34)$	& $i\sqrt{1-z}$	& $-i\sqrt{1-\bar z}$		& $i\sqrt{1-\bar z}$ 	&	$-i\sqrt{1-z}$	\\\hline
$(13)(24)$	& $\sqrt{\bar z(1-z)}$		& $\sqrt{\bar z(1-\bar z)}$		& $\sqrt{z(1-\bar z)}$ 		&	$\sqrt{z(1-z)}$			\\\hline
$(14)(23)$	& $i\sqrt{\bar z}$		& $-i\sqrt{\bar z}$ 		& $i\sqrt{z}$		&	$-i\sqrt{z}$		\\\hline
\end{tabular}
\end{center}
The trick now is that these $\tilde h_i(\pi)$ satisfy the group property
\beq
	n(\tilde h_i(\pi\sigma))=n(\tilde h_i(\pi))n(\tilde h_{\pi^{-1}(i)}(\sigma)),
\eeq
which is an identity in $Spin(2)$, while it is only trivial that it holds in $SO(2)$. 

This fact together with the fact that the action of $\mathbb{Z}_2^2$ is free actually implies that these phases can be trivialized in the following way. Suppose for concreteness that the full symmetry is the $\mathbb{Z}_2^2$, the argument for subgroups is similar. Thus, assume that all polarizations $\pol_i$ transform in the same representation $\rho$ and denote by $\rho(h)$ the action of $n(h)\in Spin(2)$. First, define the new polarizations
\al{
	\tilde \pol_1&=\pol_1,\nn\\
	\tilde \pol_2&=\rho(-i\sqrt{1-\bar z})\pol_2,\nn\\
	\tilde \pol_3&=\rho(\sqrt{z(1-\bar z)})\pol_3,\nn\\
	\tilde \pol_4&=\rho(-i\sqrt{z})\pol_4.
}
Then recall that the action of the permutation, say, $(14)(23)$ is
\al{
	\pol_1&\to \rho(-i\sqrt{z})\pol_4,\quad
	\pol_4\to \rho(i\sqrt{\bar z})\pol_1,\nn\\
	\pol_2&\to \rho(i\sqrt{z})\pol_3,\quad
	\pol_3\to \rho(-i\sqrt{\bar z})\pol_2.
}
This induces the following action on the redefined polarizations,
\al{
	\tilde \pol_1&\to \rho(-i\sqrt{z})\pol_4=\tilde \pol_4,\nn\\
	\tilde \pol_2&\to \rho(-i\sqrt{1-\bar z})\rho(i\sqrt{z})\pol_3=\tilde \pol_3,\nn\\
	\tilde \pol_3&\to \rho(\sqrt{z(1-\bar z)})\rho(-i\sqrt{\bar z})\pol_2=\tilde \pol_2,\nn\\
	\tilde \pol_4&\to \rho(-i\sqrt{z})\rho(i\sqrt{\bar z})\pol_1=\tilde \pol_1.
}
It is easy to check that the same holds for all other permutations. Since the redefinition commutes with the action of the stabilizing $O(d-2)$, we conclude that for the purposes of counting the structures we simply look at the tensor product $\bigotimes_{i=1}^4\rho_i$ symmetrized by the kinematic symmetry group of the correlator \textit{without} the fermionic $-$ sign, and then extract the $O(d-2)$ singlets.

For completeness we also consider the non-kinematic permutations. It is sufficient to consider $(12)$ and $(13)$ since these together with the kinematic permutations generate the full $S_4$. For these permutations $x'_i\neq x_i$, but rather
\begin{center}
\begin{tabular}{r||c|c|c|c}
 & $x_1'$ & $x_2'$ & $x_3'$ & $x_4'$ \\
\hline
\hline
(12) & 0 & $z/(z-1)$ & 1 &$\infty$  \\
\hline
(13) & 0 & $1-z$ & 1 & $\infty$
\end{tabular}
\end{center}
We find the following permutation phases
\begin{center}
\begin{tabular}{r || c | c | c | c }
			& $h_1$ 		& $h_2$ 		& $h_3$ 		& $h_4$ 		\\
\hline\hline
$(12)$	& $\sqrt{1-\bar z}$	& $\sqrt{1-\bar z}$		& $\sqrt{1-\bar z}$ 	&	$\sqrt{1-z}$	\\\hline
$(13)$	& $i$		& $i$		& $i$ 		&	$-i$			\\\end{tabular}
\end{center}
Again, we can define $\tilde h$ to automatically account for fermionic ``$-$'' sign, 
\begin{center}
\begin{tabular}{r || c | c | c | c }
			& $\tilde h_1$ 		& $\tilde h_2$ 		& $\tilde h_3$ 		& $\tilde h_4$ 		\\
\hline\hline
$(12)$	& $\sqrt{1-\bar z}$	& $-\sqrt{1-\bar z}$		& $\sqrt{1-\bar z}$ 	&	$\sqrt{1-z}$	\\\hline
$(13)$	& $i$		& $i$		& $-i$ 		&	$-i$			\\\end{tabular}
\end{center}

\section{Character formula for symmetrized tensor products}
\label{app:characters}
Consider a tensor product
\beq
	W=V^{\otimes n},
\eeq
and the subspace of it invariant under a subgroup $\Pi\subseteq S_n$ of permutations of tensor factors,
\beq
	Z=\left[V^{\otimes n}\right]^\Pi.
\eeq
More generally, we can allow $\Pi$ to act by multiplication by permutations followed by a multiplication by a one-dimensional character $\chi_\Pi$ of $\Pi$. As an example, we can have $\Pi=S_n$ and $\chi_\Pi(\pi)\equiv 1$, in which case $Z$ is the $n$-th symmetric tensor power, or $\chi_\Pi(\pi)=\mathrm{sign}\,\pi$, in which case $Z$ is the $n$-th antisymmetric power of $V$. For simplicity, we will consider only these two choices of $\chi_\Pi$, but leave $\Pi$ completely general.

Assume that $V$ is a representation of some group $G$, given by $\rho:\,G\to GL(V)$. Then $W$ and $Z$ are also representations of $G$, and out goal is to compute the character of $G$ on $Z$, $\chi_Z$.

Define the operator
\beq
	P=\frac{1}{|\Pi|}\sum_{\pi\in\Pi}\pi\in GL(W),
\eeq
where $\pi$ acts as described above. Let $\rho_n=\rho^{\otimes n}$. Note that
\beq
	P^2=\frac{1}{|\Pi|^2}\sum_{\pi,\sigma\in\Pi}\pi\sigma=\frac{1}{|\Pi|^2}\sum_{\pi,\sigma'\in\Pi}\pi\pi^{-1}\sigma'=P,
\eeq
where $\sigma'=\pi\sigma$. Since $P^2=P$, $P$ is a projection and $W$ decomposes into a sum of eigenspaces of $P$, $W=W_0\oplus W_1$, with the explicit decomposition being $w=(1-P)w+Pw$. It is easy to see that $Pw$ is $\Pi$-invariant and if $w$ is $\Pi$-invariant, then $Pw=w$. This shows $W_1=Z$. Since $\rho_n$ commutes with $P$, this decomposition is also a decomposition of $W$ into representations of $G$. It then follows that
\beq
	\chi_Z(g)=\mathrm{tr}\,P\rho_n(g),
\eeq
as can be shown by choosing a basis diagonal for $P$. It is a simple exercise to show in some choice of basis that
\beq
\label{eq:generalcharacter}
	\chi_Z(g)=\frac{1}{|\Pi|}\sum_{c\in C}\left[|c|\chi_\Pi(c)\prod_i \chi_\rho(g^{c_i})\right],
\eeq
where $C$ is the set of cycle types of permutations in $\Pi$, $|c|$ is the number of elements of cycle type $c$ in $\Pi$, and $c_i$ are the cycle lengths in the cycle type $c$. For example, the cycle type of the trivial permutation is $c=1^n$, i.e.\ it is a product of $n$ cycles of length $1$, and $|c|=1$. Therefore the contribution of the identity to the sum is always $\chi_\rho^n(g)$. Since we restricted $\chi_\Pi$ to come from a one-dimensional character of $S_n$, it takes the same value on all elements with the same cycle type, so that notation $\chi_\Pi(c)$ is well-defined.

The examples relevant in this paper are $\Pi=\mathbb{Z}_2\subset{S_2}$, $\Pi=S_3$ and $\Pi=\mathbb{Z}_2^2\subset{S_4}$. In the first case we have two cycle types, $1^2$ and $2^1$, each occuring once, and therefore we obtain for the trivial $\chi_\Pi$
\beq
	\chi_{\mathbb{Z}_2}(g)=\mathrm{S}^2\chi(g)=\frac{1}{2}\left[\chi^2(g)+\chi(g^2)\right],\label{eq:symmetriccharacter}
\eeq
the well-known formula for the symmetric square. For the exterior square one has, using $\chi_\Pi=\mathrm{sign}$,
\beq
	\wedge^2\chi(g)=\frac{1}{2}\left[\chi^2(g)-\chi(g^2)\right].\label{eq:wedgecharacter}
\eeq

In the second case we have the symmetric and exterior cube relevant for proposition~\ref{prop:sthree}. In this case we have $\Pi=S_3$ and cycle types $1^3,2^21^1,3^1$ with multiplicities $1,3,2$. We find from~\eqref{eq:generalcharacter},
\al{
	\mathrm{S}^3\chi(g)&=\frac{1}{6}\left[\chi(g)^3+3\chi(g^2)\chi(g)+2\chi(g^3)\right],\\
	\wedge^3\chi(g)&=\frac{1}{6}\left[\chi(g)^3-3\chi(g^2)\chi(g)+2\chi(g^3)\right].
}

In the third case we have cycle types $1^4$ and $2^2$ with the latter occuring thrice, so that we find
\beq	
\label{eq:characterz2z2}
	\chi_{\mathbb{Z}_2^2}(g)=\frac{1}{4}\left[\chi^4(g)+3\chi^2(g^2)\right].
\eeq
In practice this can be computed as
\beq
	\rho^4\ominus3\p{\wedge^2\rho\otimes \mathrm{S}^2\rho },
\eeq
which easily can be checked using the above formulas. The case $\chi_\Pi=\mathrm{sign}$ is equivalent to $\chi_\Pi\equiv 1$.

\bibliographystyle{JHEP}
\bibliography{conformal_frame}

\end{document}